\documentclass[superscriptaddress,longbibliography,amsmath,amssymb,aps,prx,notitlepage,twocolumn,floatfix]{revtex4-1}

\usepackage{graphicx,xcolor}
\usepackage{amsmath,amssymb}
\usepackage[T1]{fontenc}
\usepackage[ngerman, english]{babel}
\usepackage{hyphenat}
\usepackage{dsfont}
\usepackage{bm}
\usepackage{color}
\usepackage{lipsum}
\usepackage{MnSymbol, marvosym, wasysym}

\usepackage{microtype}

\usepackage[hidelinks]{hyperref}
\hypersetup{
    linkcolor=blue,
    colorlinks=true,
    citecolor=blue
    }
\usepackage{url}

\begin{document}
\renewcommand{\vec}[1]{\boldsymbol{#1}}
\newcommand{\up}{{\uparrow}}
\newcommand{\dw}{{\downarrow}}
\newcommand{\pa}{{\partial}}
\newcommand{\pd}{{\phantom{\dagger}}}
\newcommand{\bs}[1]{\boldsymbol{#1}}
\newcommand{\todo}[1]{{\textbf{\color{red}#1}}}
\newcommand{\q}[1]{{\textbf{\color{blue}#1}}}
\newcommand{\g}[1]{{\textbf{\color{green}#1}}}
\newcommand{\sr}[1]{{\color{red}#1}}
\newcommand{\srb}[1]{{\color{blue}#1}}
\newcommand{\dm}[1]{{\color{red}#1}}
\newcommand{\eps}{{\varepsilon}}
\newcommand{\nn}{\nonumber}
\newcommand{\ie}{{\it i.e.},\ }
\def\eg{\emph{e.g.}\ }
\def\ea{\emph{et al.}}
\def\cf{\emph{c.f.}\ }

\title{Quasiparticle excitations in a one-dimensional interacting topological insulator: Application for dopant-based quantum simulation}

\author{David Mikhail}
\affiliation{School of Physics, University of Melbourne, Parkville, VIC 3010, Australia}
\author{Benoit Voisin}
\affiliation{Silicon Quantum Computing, Sydney, NSW 2052, Australia}
\affiliation{Centre for Quantum Computation and Communication Technology, School of Physics, The University of New South Wales, Sydney, NSW 2052, Australia}
\author{Dominique Didier St Medar}
\affiliation{Silicon Quantum Computing, Sydney, NSW 2052, Australia}
\affiliation{Centre for Quantum Computation and Communication Technology, School of Physics, The University of New South Wales, Sydney, NSW 2052, Australia}
\author{Gilles Buchs}
\altaffiliation[Current address: ]{Quantum Information Science Section, Oak Ridge National Laboratory, Oak Ridge, TN 37831, USA}
\affiliation{Silicon Quantum Computing, Sydney, NSW 2052, Australia}
%\affiliation{Quantum Information Science Section, Oak Ridge National Laboratory, Oak Ridge, TN 37831, USA}
\author{Sven Rogge}
\affiliation{Centre for Quantum Computation and Communication Technology, School of Physics, The University of New South Wales, Sydney, NSW 2052, Australia}
\author{Stephan Rachel}
\altaffiliation[Corresponding author: ]{{\tt stephan.rachel@unimelb.edu.au}}
%\email[Corresponding author: ]{\tt stephan.rachel@unimelb.edu.au}
\affiliation{School of Physics, University of Melbourne, Parkville, VIC 3010, Australia}

\date{\today}

\begin{abstract}
We study the effects of electron-electron interactions on the charge excitation spectrum of the spinful Su-Schrieffer-Heeger (SSH) model, a prototype of a one-dimensional bulk obstructed topological insulator.
In view of recent progress in the fabrication of dopant-based quantum simulators 
we focus on experimentally detectable signatures of interacting topology in finite lattices.
To this end we use Lanczos-based exact diagonalization to calculate the 
single-particle spectral function in real space which generalizes the local density of states to interacting systems. Its spatial and spectral resolution allows for the direct investigation and identification of edge states.
By studying the non-interacting limit, we demonstrate that the topological in-gap states on the boundary are robust against both finite-size effects as well as random bond and onsite disorder which suggests the feasibility of simulating the SSH model in engineered dopant arrays in silicon.
While edge excitations become zero-energy spinlike for any finite interaction strength, 
our analysis of the spectral function shows that 
the single-particle charge excitations are gapped out on the boundary.
Despite the loss of topological protection we find that these edge excitations are quasiparticle-like as long as they remain within the bulk gap. Above a critical interaction strength of $U_c\approx 5 t$ these quasiparticles on the boundary lose their coherence which is explained by the merging of edge and bulk states. 
This is in contrast to the many-body edge excitations which survive the limit of strong coupling, as established in the literature.
Our findings show that for moderate repulsive interactions the non-trivial phase of the interacting SSH model can be detected through remnant signatures of topological single-particle states using single-particle local measurement techniques such as scanning tunneling spectroscopy.
\end{abstract}

\maketitle
%\begin{textblock}\noindent\fontsize{7}{7}\selectfont\textcolor{black!30}{Notice: This manuscript has been co-authored by UT-Battelle, LLC, under contract DE-AC05-00OR22725 with the US Department of Energy (DOE). The US government retains and the publisher, by accepting the article for publication, acknowledges that the US government retains a nonexclusive, paid-up, irrevocable, worldwide license to publish or reproduce the published form of this manuscript, or allow others to do so, for US government purposes. DOE will provide public access to these results of federally sponsored research in accordance with the DOE Public Access Plan (http://energy.gov/downloads/doe-public-access-plan).}\end{textblock}

%%%%%%%%%%%%%%%%%%%%%%%%%%%%%%%%%%%%%%%%%%%%%%%%%%%%%%%%%%%%%%
%
%                   I N T R O D U C T I O N
%
%%%%%%%%%%%%%%%%%%%%%%%%%%%%%%%%%%%%%%%%%%%%%%%%%%%%%%%%%%%%%%

\section{Introduction\label{sec::introduction}}
The research of topological phases of matter in condensed matter physics has rapidly grown since the theoretical prediction\,\cite{Kane2005a,Kane2005b,Bernevig2006,Bernevig2006a}
and experimental discovery\,\cite{Koenig2007} of topological insulators (TIs).
A TI is characterized by an insulating bulk with conducting zero-energy states at its boundary. 
These boundary states are protected against local perturbations as long as certain symmetries are preserved and the bulk gap does not close. 
For non-interacting TIs, all possible topological phases, distinguished by a $\mathbb{Z}$ or $\mathbb{Z}_2$ number called topological invariant, are classified in dependence of the spatial dimension and anti-unitary symmetries of the single-particle  Hamiltonian\,\cite{Schnyder2008a,Kitaev2009b}. This classification-scheme does not apply to many-body Hamiltonians and, except for fermions in 1D\,\cite{Pollmann2010a,Chen2011}, the classification of interacting topological phases remains an open problem.  
As demonstrated for many systems\,\cite{Rachel2018a,Hohenadler2013} the effects of electron correlations can be versatile. In 1D quartic interactions can modify the classification of topological phases\,\cite{Fidkowski2010c,Fidkowski2011} while in the case of the proposed topological Mott insulator, repulsive Hubbard interactions can generate non-trivial topological states\,\cite{Raghu2008e}.
The study of such strongly-correlated electron materials, typically framed in terms of variants of the Fermi-Hubbard model, is notoriously difficult as these systems are neither amenable to perturbative methods nor to classical numerical simulations.  
The large array of exotic phenomena originating from the presence of strong electron electron interactions, ranging from high-$T_c$ superconductors and spin liquids to correlated TIs, have fueled efforts to realize fermionic many-body physics in quantum simulation platforms like ultra cold atoms\,\cite{Bloch2012,Georgescu2014} and semiconductor nanostructures\,\cite{Hensgens2017i}. In particular, recent advances in the engineering of dopant-based quantum dot arrays using the atomic precision of scanning tunneling lithography have brought the simulation of Fermi-Hubbard Hamiltonians in solid state systems within feasible reach\,\cite{Wyrick2019,Wang2021,Kiczynski2022}. A large range of parameters can be accessed in dopant systems to observe the desired fermionic phases, including strong Hubbard interactions and low effective temperatures\,\cite{Salfi2016}.

One important advantage of these semiconductor architectures over other platforms is the combination of standard non-local transport measurements\,\cite{Kiczynski2022} with single-site measuring techniques like scanning tunneling spectroscopy (STS), a local method which reveals real space information (although measured high-quality data can in principle also provide information in momentum space via Fourier transformation\,\cite{Schneider,Buchs2009}). It allows to probe electronic energy spectra as well as local spin and charge transport\,\cite{Salfi2016, Voisin2020} and has proven to play a pivotal role for the study of topological band structures\,\cite{Roushan2009,Alpichshev2010,Reis2017,Schneider}.
STS measures the tunneling conductivity $dI/dV$ for an applied voltage $V$ which is proportional to the local density of states (LDOS) of the sample surface. 
Lately, STS has repeatedly been applied to correlated systems\,\cite{Pan2001,Becker2011,Adler2019,Wu2020} and there are first attempts to understand $dI/dV$-spectra in the presence of strong electron interactions\,\cite{Rontani2005a,Maruccio2007,Secchi2012a,Ervasti2017,Varney2010h};
$dI/dV$ can be formalized in terms of the single-particle spectral function (SPSF) in real space\,\cite{Ervasti2017} which generalizes the LDOS to interacting systems.
The goal of this work is to demonstrate the utility of the real space SPSF as a tool to understand STS spectra in the presence of electron correlations and in particular to identify interacting topological phases.   

To this end we calculate the SPSF of a correlated topological insulator in 1D with an emphasis on quasiparticle edge-excitations. As a minimal model combining both strong correlations and non-trivial topology we study the Su-Schrieffer-Heeger-Hubbard (SSHH) model of spin-$1/2$ fermions, sometimes also referred to as Peierls-Hubbard model\,\cite{Jeckelmann2002}.
The non-interacting Su-Schrieffer-Heeger (SSH) model\,\cite{Su1979b} describes fermions on a chain with a gapped spectrum due to an alternating hopping amplitude. At the interface of different dimerization patterns it features a bulk-gap closing and zero-energy states protected by chiral symmetry. As such it constitutes a prototype of a 1D TI of the BDI class\,\cite{Ryu,Shen2012,Kane2013,Wang2015f,Asboth2016a}, which has been realized in classical RLC circuits\,\cite{Lee2018}. In contrast to topological systems in $d>1$ the non-trivial phase admits an atomic limit which classifies it as a bulk obstructed phase\,\cite{Bradlyn2017a,Khalaf2021} (see section ``Discussion'' for details).
Different variants of the interacting SSH model were studied in the literature\,\cite{Kiczynski2022,Manmana2012,Yoshida2014,Wang2015f,Ye2016,Sirker2014a,Barbiero2018c,Sbierski2018c,Le2020,Li2022}. 
While in the spinless case strong nearest-neighbor interactions drive the system into a trivial charge density wave\,\cite{Sirker2014a}, it was shown that for spinful fermions the topological phase is robust towards repulsive onsite interactions\,\cite{Manmana2012}. There the bulk-boundary correspondence remains valid and the associated zero-energy states are found to be collective excitations of two entangled edge-spins of opposite sign\,\cite{Manmana2012,Yoshida2014,Barbiero2018c}. 

In this work, we employ Lanczos-based exact diagonalization (ED) to obtain the many-body spectrum and states from which the spectral functions are calculated. We find that the quasiparticle charge edge states lose their topological nature as a Mott gap opens up in both bulk and boundary\,\cite{Yoshida2014}. Nevertheless, we show that for a substantial range of interaction strengths distinct quasiparticle excitations remain as remnant signatures of the topological SSH edge states and serve as indicators of the underlying bulk-topology.
Ultimately, for strong couplings $U/t \gtrsim 5$ (for 12-site chains) the quasiparticle description on the boundary breaks down as edge and bulk excitations start to substantially overlap.
The rest of this paper is structured as follows: In section \ref{sec::methods} we define the SPSF and discuss the principle behind ED. The model Hamiltonians are introduced in section \ref{sec::model}. 
In section \ref{sec::free} we discuss the SPSF in the non-interacting case and present a disorder analysis.
The interacting case is presented in section \ref{sec::interacting} where we show energy spectra, SPSF and the spin-spin correlation function of the SSHH model. In section \ref{sec::odd_chains} we address the topic of odd chains.
In the discussion in section \ref{sec::discussion} we outline the consequences for future experiments and elaborate on bulk obstructed vs. topological phases. We also discuss the topological phase diagram of the SSHH model and comment on its relation to the Haldane phase of spin-1 chains.
In the last section \ref{sec::conclusion} we briefly summarize our main findings and its implications.
The principles of ED and the Lanczos algorithm are elaborated in Appendix\,\ref{app:ED}. In Appendix\,\ref{app:SPSF} we provide more details on the concept and calculation of the SPSF.

%%%%%%%%%%%%%%%%%%%%%%%%%%%%%%%%%%%%%%%%%%%%%%%%%%%%%%%%%%%%%%
%
%                   M E T H O D S
%
%%%%%%%%%%%%%%%%%%%%%%%%%%%%%%%%%%%%%%%%%%%%%%%%%%%%%%%%%%%%%%

\section{Methods \label{sec::methods}}
Owing to its ability to measure the electronic spectrum of surface states with a high spatial resolution, STS shows great utility for the investigation of topological boundary states.
The main observable in STS is the tunneling current $I$ between states of the tip and the sample's surface in dependence of the applied voltage $V$. Based on Bardeen's transfer Hamiltonian formalism\,\cite{Bardeen1961} it can be expressed as
\begin{align}
    I &=\frac{2\pi e}{\hbar} \sum_{\mu,\nu}  f(E_\mu)[1-f(E_\nu+ eV)] \nonumber\\ 
    &\times|M_{\mu\nu}|^2 \delta(E_\mu-E_\nu) \label{current1},
\end{align}
where $f(E)$ is the Fermi-function and $M_{\mu\nu}$ is the tunneling matrix element between the single-particle states of the tip $\psi_\mu$ and the surface of the sample $\psi_\nu$.

Under the right circumstances\,\cite{Wiesendanger1994}, one can show that the conductivity
\begin{align}
    \frac{\partial I}{\partial V} \propto \rho_s(r_0,E_F-eV)  \label{conductivity}
\end{align}
mainly probes variations in the LDOS of the sample's surface. This interpretation of $dI/dV$-spectra is limited to non-interacting systems since the concept of an LDOS is based on a single-particle picture. 
Assuming a current small enough for the sample to return to its ground state after each tunneling event, the LDOS\,\cite{Odashima2017} and thereby STS-tunneling\,\cite{Bruus1983} can be generalized to arbitrary interactions via the SPSF in real-space.
Formally, the real-space spectral function is defined via the imaginary part of the retarded single-particle Green's function\,\cite{Odashima2017}, i.e.,
\begin{align}
    A_{i}(\omega)= -\frac{1}{\pi} \text{Im} (G^\text{r}_{i}(\omega))\label{SF}.
\end{align}
To evaluate the Green's function we consider the numerically convenient Lehmann representation as well its continued fraction expansion. Both approaches are explained in detail in Appendix\,\ref{app:SPSF}.
We apply ED to compute the SPSF, a numerical method for solving the many-body Schr\"odinger equation which involves no approximation and produces the exact eigenenergies and eigenstates of the full interacting Hamiltonian. As such it allows to study static and dynamic correlation functions of the ground state and, in principle, for finite temperatures. It further serves to benchmark the predictions of other methods. The drawback of this method is that computational time and memory requirements are proportional to the dimension of the Hilbert space which grows exponentially. Here we restrict our analysis to sufficiently short chains, what can be dealt with using ED. This is justified by assuming that the first generation of experiments will also be limited to short chains\,\cite{Kiczynski2022}. A detailed discussion of ED can be found in Appendix\,\ref{app:ED}.

%%%%%%%%%%%%%%%%%%%%%%%%%%%%%%%%%%%%%%%%%%%%%%%%%%%%%%%%%%%%%%
%
%                   M O D E L
%
%%%%%%%%%%%%%%%%%%%%%%%%%%%%%%%%%%%%%%%%%%%%%%%%%%%%%%%%%%%%%%

\section{Model \label{sec::model}}
To investigate the interplay of topological band structure and repulsive electron-correlations in one dimension (1D), we study the low-energy physics of the SSHH model, a dimerized variant of the Fermi-Hubbard model. The kinetic properties are described by the tight-binding SSH Hamiltonian, which allows for a topological phase transition while interactions are taken into account by a local Hubbard-term. 
Before introducing the SSHH model we begin with a brief presentation of the basic properties of the non-interacting SSH model. Studying the free fermion case allows us to understand the effect of repulsive interactions on the topological phase.
Furthermore, since the quadratic Hamiltonian of the SSH model is exactly solvable we can verify that the relevant signatures of topological edge states appearing in large chains remain present for the limited system sizes accessible to a treatment with ED. 
\begin{figure}[t!]
\begin{center}
    \centering
         \includegraphics[width=\columnwidth]{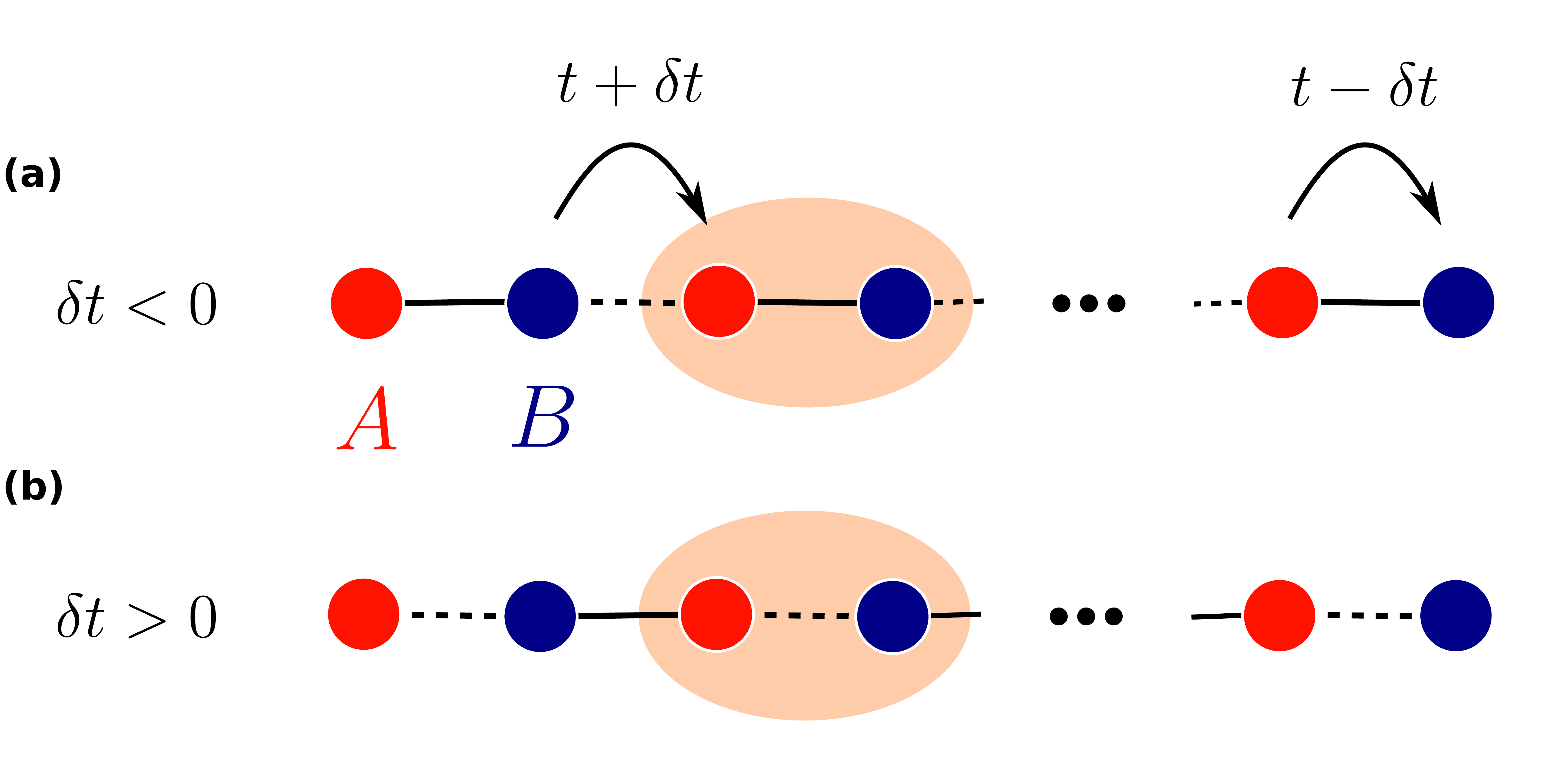}
\caption{The two different chain configurations of the SSH model. The sublattices $A$ and $B$ are coloured as red and blue sites, respectively. (a) The trivial insulating phase is characterized by a negative dimerization $\delta t<0$ such that the strong bonds (solid lines) connect two sites of the same unit cell (orange shaded area). (b) For positive dimerization $\delta t>0$ the chain becomes topological with weakly bound (dashed lines) edge sites. The strong bonds are now between sites of different unit cells. }
\label{fig:ssh}
\end{center}
\end{figure}

\subsection{Su-Schrieffer-Heeger model\label{sec::ssh model}}
\begin{figure}[t!]
\begin{center}
    \centering
         \includegraphics[width=\columnwidth]{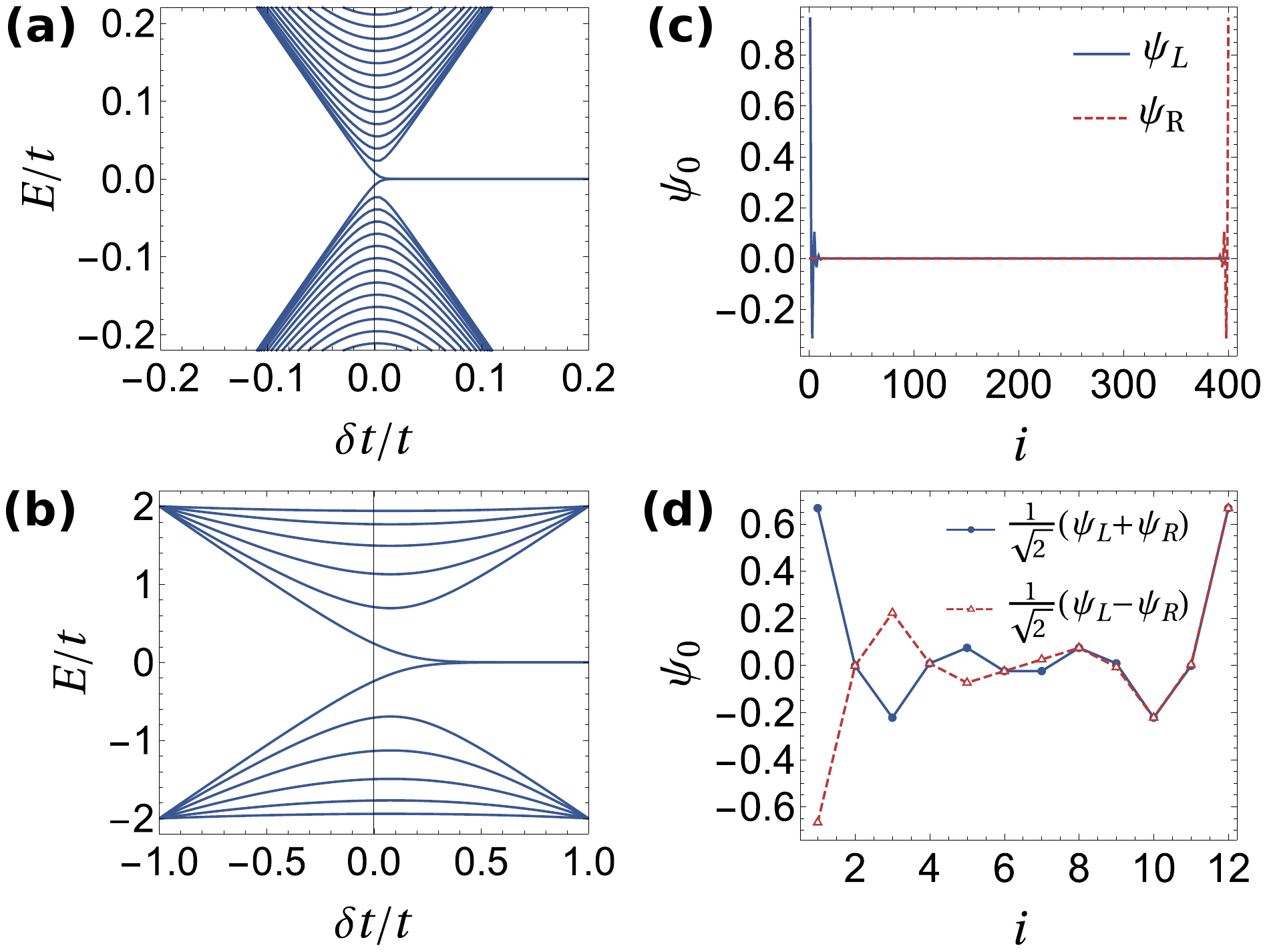}
\caption{
Single-particle spectrum and edge state wave functions of the SSH model for chains of $L=400$ ((a) and (c)) as well as $L=12$ ((b) and (d)) lattice sites. The wave functions are plotted for $\delta t/t =0.5$.}
\label{fig:finite_size_spectrum}
\end{center}
\end{figure}

The SSH model describes non-interacting spin-1/2 fermions on a one-dimensional lattice subjected to a Peierls instability which is manifested by spatially-alternating hopping amplitudes\,\cite{Su1979b}. The corresponding single-particle Hamiltonian reads
\begin{align}
    \hat{H}_{\text{SSH}}= \sum_{i,\sigma=\uparrow,\downarrow} \left(t+(-1)^i \delta t \right) \left[\hat{c}^\dagger_{i+1,\sigma}\hat{c}_{i,\sigma} + \text{H.c.}\right],
    \label{h_ssh}
\end{align}
with hopping amplitude $t$ modulated by the dimerization $\delta t$. The operator $\hat{c}_{i,\sigma}$ annihilates an electron with spin $\sigma$ on site $i$. For convenience, we choose $t\equiv 1$ which sets the energy scale for all parameters appearing in the Hamiltonian.
A staggering in the hopping amplitude leads to a doubling in size of the unit cell, which is comprised of two atoms belonging to different sublattices $A$ and $B$ depicted in Fig.\,\ref{fig:ssh}. The bipartite lattice structure implies that the Hamiltonian exhibits a chiral symmetry defined as
\begin{align}
    \hat{\Gamma}\hat{H}\hat{\Gamma}^\dagger=-\hat{H}, \label{chiral_symmetry}
\end{align} 
where $\hat{\Gamma}=\hat{P}_A-\hat{P}_B$ is a unitary and Hermitian operator representing a chiral transformation which is related to the projection operators $\hat{P}_{A}$ and $\hat{P}_{B}$ onto the sublattices $A$ and $B$, respectively. From Eq.\,\eqref{chiral_symmetry}, it follows that each eigenstate $\psi=\left(\psi_A, \psi_B \right)^\text{T}$ with energy $E$ has a chiral symmetric partner $\psi'=\hat{\Gamma}\psi=\left(\psi_A, -\psi_B \right)^\text{T}$ with energy $-E$\,\cite{Asboth2016a}.
At half filling, where the number of electrons equals the number of lattice sites, the deviation from a regular chain with uniform hoppings opens up an energy gap $\Delta E = 4|\delta t|$, rendering the system electronically insulating. 
The modulation pattern of hopping amplitudes can be thought of as alternating weak and strong bonds and the mapping $\delta t\to - \delta t$ effectively corresponds to a global translation of the bond pattern by half a unit cell which is not a symmetry of the SSH model \eqref{h_ssh}. In the case of a chain with open boundary conditions (OBC) where translational invariance is broken, this amounts to changing the nature of the edge bonds from strong to weak. The physical implications of this transformation on the edge states is best illustrated in the fully dimerized limit $|\delta t|= t$ where the weak bonds (dashed lines in Fig.\,\ref{fig:ssh}) completely vanish. For $\delta t<0$ the chain separates into strongly coupled dimers where both dimer sites are part of the same unit cell (Fig.\,\ref{fig:ssh}(a)). In contrast, as shown in Fig.\,\ref{fig:ssh}(b), for $\delta t>0$ each dimer site lies in a different unit cell which results in a chain with isolated sites on both edges, each on a different sublattice. Since hopping on these sites is completely suppressed (in the fully dimerized limit) the corresponding single-particle edge states are localized and have zero energy [Fig.\,\ref{fig:finite_size_spectrum}(a)]. 
The latter is a consequence of the chiral symmetry Eq.\,\eqref{chiral_symmetry} and the explicit representation of $\hat{\Gamma}$ in terms of sublattice operators which implies that a state with support on a single sublattice is its own chiral partner with $E=0$.
In the case of a partially dimerized chain $0<\delta t/t<1$, the edge sites are weakly coupled to the bulk of the chain and the edge-wave functions hybridize to form a chiral pair of symmetric and anti-symmetric superpositions [Fig.\,\ref{fig:finite_size_spectrum}(d)], each with finite energies of opposite sign [Fig.\,\ref{fig:finite_size_spectrum}(b)]. The hybridization of the edge states is a finite-size effect since their wave functions decay exponentially into the bulk with the system size. Hence the overlap between both edge states only vanishes in the thermodynamic limit where their energies become exactly zero, exemplified in Fig.\,\ref{fig:finite_size_spectrum}(c) where for $L=400$ sites and $\delta t/t=0.5$ the thermodynamic limit has effectively been reached. Nevertheless, even for finite systems the edge states remain highly localized, since their energies lie deep within the bulk gap close to zero energy, and only have spectral weight on either sublattice. As such, these states can be considered topological as they remain protected by chiral symmetry.

\subsection{Su-Schrieffer-Heeger-Hubbard model  \label{sec::sshh model}}

To study the effect of electronic correlations on the topological edge states of the SSH model, we complement Eq.\,\eqref{h_ssh} with repulsive onsite interactions. The full many-body Hamiltonian 
is given by 
\begingroup
\addtolength{\jot}{1em}
\begin{align}
    \hat{H}_{\text{SSHH}}&= \hat{H}_{\text{SSH}}+\hat{H}_{U} \label{total_hamiltonian}\\
    \hat{H}_U&= U\sum_i \left(\hat{n}_{i,\uparrow}-1/2\right)\left(\hat{n}_{i\downarrow}-1/2\right), \label{hubbard_hamiltonian}
\end{align}
\endgroup
where $\hat{n}_{i,\sigma}=\hat{c}^\dagger_{i,\sigma}\hat{c}_{i,\sigma}$ is the fermionic number operator and $U>0$ is the amplitude of the onsite Hubbard interaction. In the following we consider a chain of zero total magnetization, i.e.,  $N_\uparrow = N_\downarrow$. Further we focus on half filling $N=L$, where the chemical potential is $\mu=0$ and the average occupation per site is $\langle \hat{n}_{i}\rangle  = 1$. In this case Eq.\,\eqref{hubbard_hamiltonian} preserves particle-hole and time-reversal symmetry and it follows that the SSHH Hamiltonian Eq.\,\eqref{total_hamiltonian} inherits the chiral symmetry of its non-interacting counterpart\,\cite{Manmana2012}. Note that Eq.\,\eqref{hubbard_hamiltonian} is equivalent to $U\sum_i \hat{n}_{i,\uparrow} \hat{n}_{i,\downarrow}$ up to a shift of the chemical potential $\mu \to \mu + U/2$. Writing the interaction in the form of Eq.\eqref{hubbard_hamiltonian} the relation between repulsive ($U>0$) and attractive ($U<0$) interactions becomes explicit at half filling. Under the particle-hole transformation of a single spin species, say $\sigma=\,\downarrow$, given by $\hat{c}_{i,\downarrow}\to (-1)^{(i+1)} \hat{c}^\dagger_{i,\downarrow}$ with $(-1)^{(i+1)} = \pm1$ for a site on sublattice $A$ and $B$, respectively, the interaction in Eq.\,\eqref{hubbard_hamiltonian} changes sign $U\to -U$ while the hopping Hamiltonian $\hat{H}_{\text{SSH}}$ remains unaffected\,\cite{Baeriswyl1995}. Since the particle number is mapped to the magnetization (up to a constant), i.e., ($n_{i,\uparrow}-n_{i,\downarrow})/2\to 2(n_{i,\uparrow}+n_{i,\downarrow}) -1$, spin ordering in the repulsive case is related to charge ordering in the attractive case. At half filling, the energy of charge excitations, which appear in the retarded Green's function Eq.\,\eqref{GF0}, are invariant under a sign change in $U$. Hence our results for the SPSF carry over to the attractive Hubbard model (as we have tested explicitly). Similarly, the topological invariant, which can be calculated from the Green's function, remains unchanged\,\cite{Manmana2012}.

For vanishing dimerization $\delta t=0$, i.e., the gapless transition point of the SSH model, Eq.\,\eqref{total_hamiltonian} reduces to the Fermi-Hubbard Hamiltonian. It is well known that at half filling and $U>0$, this model describes a Mott insulator with an energy gap in the charge sector, i.e., adding an electron to the system inflicts an energy cost proportional to $U/t$. The insulating behavior is caused by electron localization due to the repulsive Hubbard interactions which tend to suppress double occupation such that a configuration where each lattice site is occupied by a single electron minimizes the total energy. 
For strong couplings $U/t\gg1$, second order perturbation theory shows that the dynamics of the Hubbard model are effectively described by a spin-1/2 Heisenberg Hamiltonian, i.e., 
the low energy physics are essentially governed by gapless spin-excitations\,\cite{Bethe1931,Auerbach1994}. Hence the closing of the bulk gap at the critical dimerization of the SSH model $\delta t=0$, a necessary condition of a topological phase transition, remains in the presence of repulsive interactions, although now in the spin instead of the charge sector\,\cite{Manmana2012,Yoshida2014,Le2020}.

%%%%%%%%%%%%%%%%%%%%%%%%%%%%%%%%%%%%%%%%%%%%%%%%%%%%%%%%%%%%%%
%
%                   F R E E    E L E C T R O N S 
%
%%%%%%%%%%%%%%%%%%%%%%%%%%%%%%%%%%%%%%%%%%%%%%%%%%%%%%%%%%%%%%

\section{Free electrons\label{sec::free}}

\begin{figure}[t!]
\begin{center}
    \centering
         \includegraphics[width=\columnwidth]{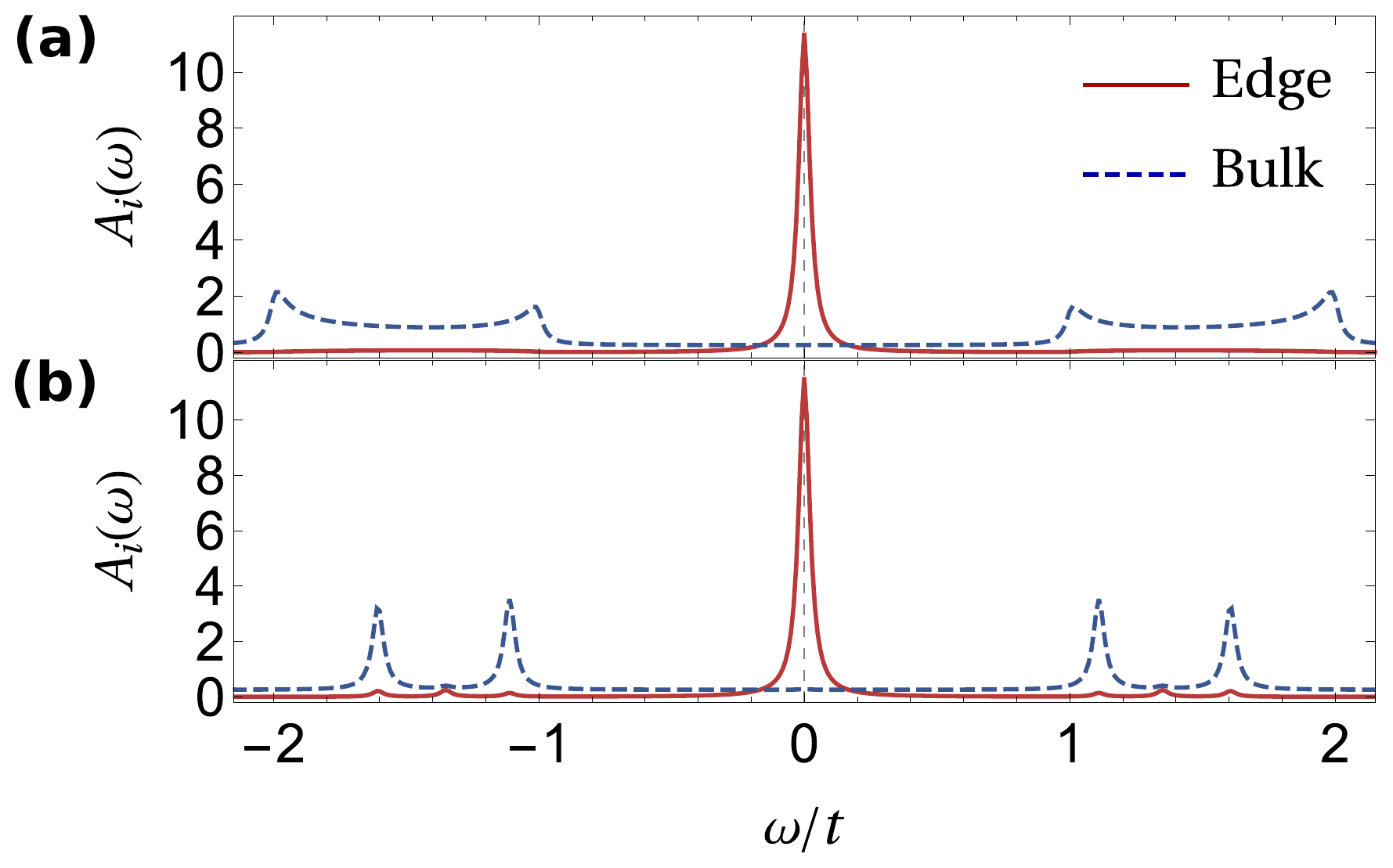}
\caption{
Single-particle spectral function of the SSH model in the topological phase ($\delta t/t =0.5 $) for (a) $L=400$ and (b) $L=12$ lattice sites. The red curves depict the edge site ($i=1$)  the blue curves refer to a bulk site ($i=L/2$). The zeroes of the latter are slightly shifted upwards for means of visibility. All results are calculated for $\eta=0.025$.}
\label{fig:finite_size_sf}
\end{center}
\end{figure}

\subsection{Finite-size analysis  \label{sec::finite size}}
For non-interacting systems the notion of a topological phase transition is based on the closing of the bulk gap and formally applies only to a system in the thermodynamic limit ($L\to \infty$) where the energy levels form dense bands.
In contrast, for a finite number of lattice sites, the energy spectrum becomes discrete which may yield a finite energy gap for all values of the dimerization. Furthermore, the dimerization for which the bulk gap reaches its minimum is shifted from zero to a positive value. Both, the size of the energy gap at its minimum as well as the shift in dimerization grow with decreasing system size which can be seen in Fig.\,\ref{fig:finite_size_spectrum} where the single-particle spectrum is shown for a chain of (a) $L=400$ and (b) $L=12$ sites. Despite the absence of a gap closing, for sufficiently large and positive dimerization strengths two eigenstates appear close to zero energy within the bulk gap corresponding to the hybridized topological edge states. Their real-space wave functions are shown in Fig.\,\ref{fig:finite_size_spectrum}(c) and (d) for both system sizes. While in the large-$L$ limit, each edge state is localized on a single edge, at finite chain-lengths the overlap between both wave functions becomes sufficiently large for them to hybridize into symmetric and anti-symmetric superpositions.
The SPSF of an edge and bulk site in the topological phase is shown in Fig.\,\ref{fig:finite_size_sf} for chains of (a) $L=400$ and (b) $L=12$ sites, respectively. The zero-line of the bulk SPSF is shifted slightly upwards to support visibility.
As expected for the insulating SSH model in the limit of large chain lengths ($L=400$), the bulk states in Fig.\,\ref{fig:finite_size_sf}(a) form two bands separated by an energy gap. For positive values of the dimerization the SPSF of an edge site features two excitation peaks within the bulk gap close to zero energy. Due to a finite peak-width and energy discretization, these peaks both appear to lie at $\omega=0$. Formally, this would only be the case either in the thermodynamic limit ($L\to \infty $ with $N/L=1$) or fully-dimerized limit ($\delta t = t$) where the overlap between the wave functions of both edges vanishes and their respective energies become exactly zero. In the limit of $\eta \to 0^+$, the zero-energy peak splits into two delta functions located at small but finite energies symmetrically around $\omega =0 $.
Since experimental measurements are conducted at finite temperatures and limited to finite energy resolutions, the observed excitation spectra are always subject to peak broadening.
To account for this deviation from a delta-peaked spectrum we chose a convergence factor of $\eta=0.025$.

The origin of the edge peaks can be understood by considering the single-particle processes which are mapped out by the SPSF.
Since at half filling only two of the four possible edge states (spin up and down per edge) are filled, adding an electron to the edge will lead to further occupation of a zero-energy state corresponding to an 
% \dm{addition} 
peak at $\omega = 0 $. Similarly, the removal of an electron from one of the zero-energy edge states does not change the overall energy.
Due to the coinciding location of 
% \dm{addition and removal peaks}
both excitations, the amplitude of the resulting peak in Fig.\,\ref{fig:finite_size_sf} is twice as large as for the individual transition processes.
For comparison, in Fig.\,\ref{fig:finite_size_sf}(b) we again show the SPSFs for an edge and a bulk site, this time for a 12-site chain. The increased level spacing of smaller systems is clearly visible in the peak-structure of the bulk states. Further the ratio of peak amplitudes between edge and bulk states is strongly affected by the number of lattice sites. In smaller systems the conserved spectral weight is distributed over fewer excitations which is compensated by an increase in the peak amplitudes.
Nevertheless, as long as the dimerization is large enough the essential signatures for determining the topological nature of the system, i.e., the bulk gap and edge excitations close to zero-energy, remain present even for small chains. 
\begin{figure}[t!]
\begin{center}
    \centering
         \includegraphics[width=\columnwidth]{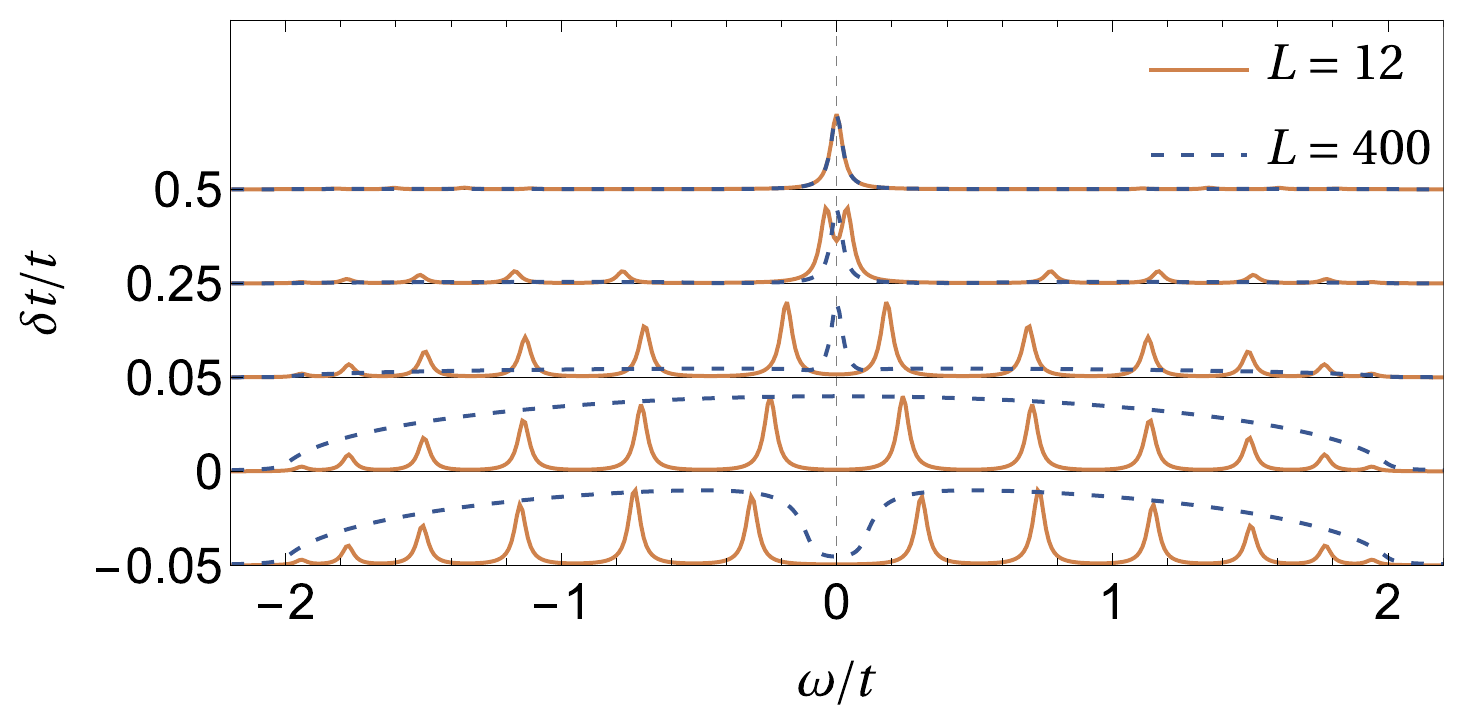}
\caption{SPSF of an edge site across the phase transition from topological ($\delta t >0 $) to trivial ($\delta t <0 $) phase. The orange curves are calculated for 12 sites while the blue ones refer to chains of 400 sites. All curves are normalized with respect to their maximum.}
\label{fig:ssh_sf_phase_transition}
\end{center}

\end{figure}
In Fig.\,\ref{fig:ssh_sf_phase_transition} we show the SPSF of an edge site across the topological phase transition for small (orange) and large (blue) chains. In the thermodynamic limit the SPSF changes drastically when crossing the critical value $\delta t=0$. The zero-energy peak persists for small but positive dimerization and vanishes when the dimerization changes its sign where edge excitations become gapped as expected for a trivial insulator. At $\delta t=0$ the bulk gap closes and the SSH model reduces to the simple tight-binding chain with metallic energy spectrum. In contrast, for small chains both edge states overlap and gap out close to the transition point $\delta t=0$, as expected from Fig.\,\ref{fig:finite_size_spectrum}(a) and only for $\delta t/t\gtrsim 0.35$ (with $\eta=0.025$) the edge states become (approximately) gapless and peaked at zero energy. On the other hand for $\delta t\leq 0 $ the same SPSF becomes bulk-like which clearly indicates a trivial insulating phase.
In conclusion, the topologically protected edge states of the SSH model remain present for the small system sizes which are accessible to the numerical treatment with ED (and small quantum dot arrays) as long as the dimerization is large enough such that the system lies deep within the topological phase. Furthermore, the relevant signatures to experimentally identify a topological phase in STS experiments, namely the zero-energy excitations on the boundary, are observable in the SPSF which in the limit of $\eta \to 0^+$ agrees with the LDOS for non-interacting systems.      

\begin{figure}[t!]
\begin{center}
    \centering
         \includegraphics[width=\columnwidth]{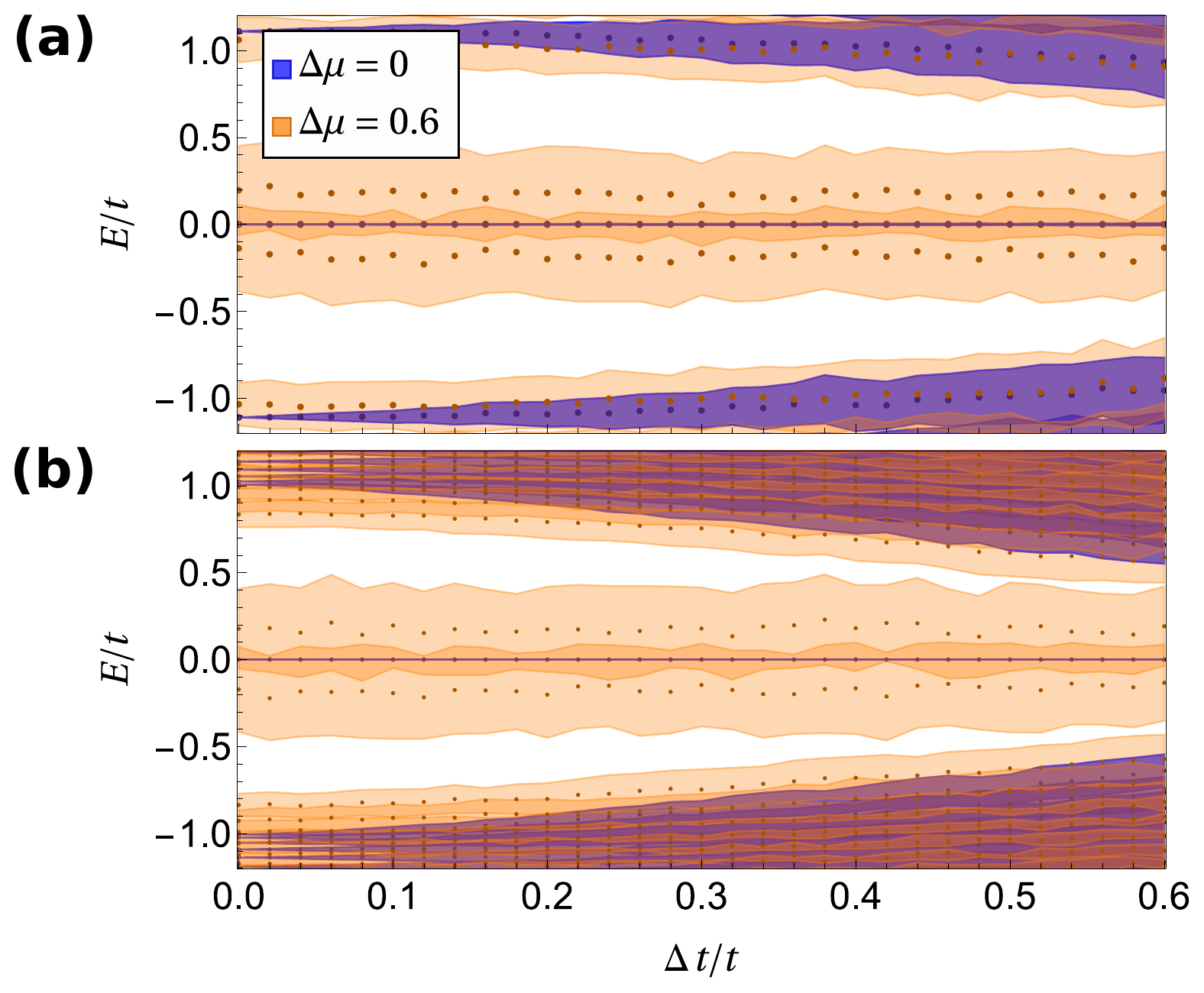}
\caption{Disordered energy spectra averaged over 150 random disorder realizations of a dimerized chain ($\delta t/t=0.5$) of 12 sites (a) and 50 sites (b), respectively. The standard deviation of the disorder averaging is depicted by blueish ($\Delta \mu=0$) and yellowish ($\Delta \mu=0.6$) shaded areas. Edge states for $\Delta \mu=0$ ($\Delta \mu\neq0$) are pinned to $E=0$ (scattered away from $E=0$).}
\label{fig:ssh_disorder}
\end{center}
\end{figure}

\subsection{Disorder analysis\label{sec::disorder}}
On the path towards scalable quantum simulation in semiconductor nanostructures, a crucial question is whether the experimental signatures are robust against lattice imperfections and other sources of disorder encountered experimentally.
In this section we discuss two common types of random disorder, namely non-magnetic bond and onsite disorder. The effect of both types on the localization of edge states in the SSH model has been studied previously in Refs.\,\cite{Perez-Gonzalez2019a,Scollon2020}.
The former refers to local variations in the hopping amplitudes which we incorporated by adding a site-dependent hopping contribution $t_i$ to Eq.\,\eqref{h_ssh}, randomly sampled from an interval $\left[-\Delta t,\Delta t\right]$.
The second disorder contribution we consider represents the influence of stray electrostatic fields. To account for these we include an onsite potential $\mu_i$ randomly drawn from an interval $\left[-\Delta \mu,\Delta \mu\right]$, which has the form of a local chemical potential.
The modified Hamiltonian is given by
\begin{align}
 \hat{H}^{\text{dis}}_{\text{SSH}}= &\sum_{i,\sigma=\uparrow,\downarrow} \left(t+(-1)^i \delta t + t_i \right) \left[\hat{c}^\dagger_{i+1,\sigma}\hat{c}_{i,\sigma} + \text{H.c.}\right]\nonumber \\&+\sum_i \mu_i \hat{n}_i, \label{disorder_hamiltonian}
\end{align}
with occupation number operator $\hat{n}_i=\hat{n}_{i,\uparrow}+\hat{n}_{i,\downarrow}$. We focus in our analysis on the impact both forms of disorder have on the spectral properties of the edge modes as those are the crucial topological signatures accessible in STS experiments. Therefore, we calculated the energy spectrum of the disordered Hamiltonian Eq.\,\eqref{disorder_hamiltonian} averaged over 150 random disorder configurations and in dependence of the disorder interval $\Delta t$. The results for a topological chain ($\delta t/t=0.5$) are shown in Fig.\,\ref{fig:ssh_disorder} with (orange) and without (blue) onsite disorder present. We calculated spectra for 12 (a) and 50 (b) sites. The dots depict the averaged single-particle energies while the shaded areas correspond to the standard-deviation due to disorder-averaging. For the edge states to be experimentally detectable in STS, their respective energies have to lie within the bulk gap. 
%\dm{For the edge states to remain localized, their respective energies have to lie within the bulk gap. }
When only bond disorder is considered, the spectrum remains gapped even for strong disorder. Since the gap-size between the bulk bands is dependent on $\delta t$ ($\Delta E=4|\delta t| $ in the thermodynamic limit) it is affected by a modulation of the dimerization. Nevertheless, as long as the global bond-pattern is unchanged the chiral symmetry remains intact and the edge-state energies remain pinned at $E=0$ as shown by the blue dots in Fig.\,\ref{fig:ssh_disorder}. 
Once random onsite potentials are included the chiral symmetry is broken and the edge states are no longer symmetry-protected. As long as the onsite potential, which constitutes a constant shift in energy, remains smaller than half of the gap-size, %\dm{, the edge states remain localized.} 
the edge states remain energetically separated from the bulk states. In Refs.\,\cite{Perez-Gonzalez2019a,Scollon2020} it was shown that despite the loss of symmetry-protection, the edge states are localized even for strong onsite disorder due to Anderson localization. Since the bulk gap shrinks with system size (cf. Fig.\,\ref{fig:finite_size_spectrum}), the effect of onsite disorder is more pronounced in larger systems. In summary, our analysis suggests that for small chains the %\dm{topological signatures} 
spectral edge signatures of the SSH model are detectable even in the presence of chiral-symmetry breaking onsite disorder. For large chains ($L\approx 50$) the in-gap states remain robust as long as the combined bond and  onsite disorders do not significantly exceed half of the gap size $\Delta E \approx 4\delta t$. We conclude that due to its large energy gap and robust edge states, the SSH model represents a promising candidate for the study of non-trivial topology in artificial dopant lattices.

%%%%%%%%%%%%%%%%%%%%%%%%%%%%%%%%%%%%%%%%%%%%%%%%%%%%%%%%%%%%%%
%
%           I N T E R A C T I N G    E L E C T R O N S 
%
%%%%%%%%%%%%%%%%%%%%%%%%%%%%%%%%%%%%%%%%%%%%%%%%%%%%%%%%%%%%%%

\section{Interacting electrons\label{sec::interacting}}
The framework of topological band theory, which relies on a single-particle picture, is no longer valid for the description of the topological properties of correlated electrons. 
That is, in the interacting case a topological phase transition does not necessarily involve the closing of the single-particle (charge) gap\,\cite{Varney2011h}, which for a many-body system at filling $N$ is defined as $\Delta E_{\text{SP}}=(E_\text{add}(N+1)-E_\text{add}(N))/2$ with addition energy $E_\text{add}(N)= E_0(N)-E_0(N-1)$. 
Instead, a change in topology can be induced by the vanishing and reopening of the many-body (spin) gap $\Delta E_{\text{MB}}=E_1-E_0$ between ground- and first excited state in the eigenenergy spectrum\,\cite{Varney2011h,Sbierski2018c}. For this reason we inspect charge and spin sectors of the SSHH model separately for signs of a topological phase transition and interacting edge states by independently varying the dimerization $\delta t/t$ and Hubbard interaction $U/t$, respectively. In contrast to a system in the thermodynamic limit, the energy levels of a finite chain are discrete and instead of a gap-closing, a phase transition is signalled by a local minimum of the bulk gap at periodic boundary conditions (PBC). 

\begin{figure}[t!]
\begin{center}
    \centering
         \includegraphics[width=\linewidth]{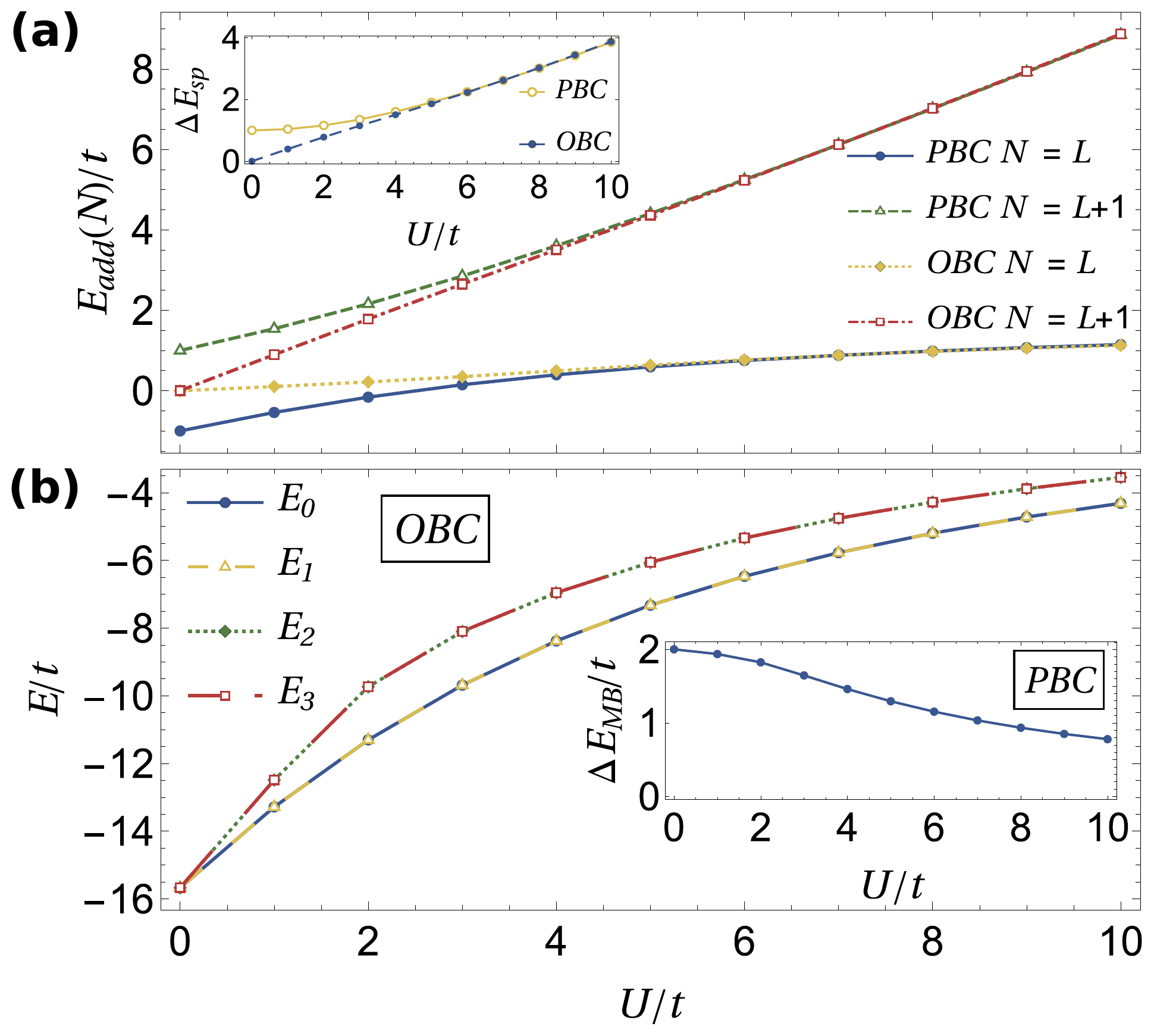}
\caption{Energy spectra of a chain of $L=12$ sites calculated for dimerization $\delta t/t=0.5$ at zero total magnetization ($S_z=0$). (a) Addition energies $E_\text{add}(N) = E_0(N+1) -E_0(N)$ of single-particle excitations between fillings $L-1 \to L $ and $L \to L+1 $ for PBC (blue and green) and OBC (yellow and red). Inset: For $U/t<U_c/t\approx5$ the open chain exhibits in-gap excitations on the edges. For $U/t\gtrsim U_c/t$ the open chain becomes bulk-like indicated by the alignment of the single-particle (charge) gap for PBC and OBC. (b) The four lowest many-body eigenenergies of an open chain. At $U=0$, the ground state is four-fold degenerate up to a small finite-size splitting (see text). For $U>0$ the degeneracy is reduced as two eigenstates with doubly-occupied edges increase in energy. Inset: Many-body (spin) gap $\Delta E_{\text{MB}}=E_1-E_0$ in the bulk between ground- and first excited eigenstate at filling $N=L$. 
}
\label{spec_U}
\end{center}
\end{figure}

\begin{figure}[h!]
\begin{center}
    \centering
         \includegraphics[width=\linewidth]{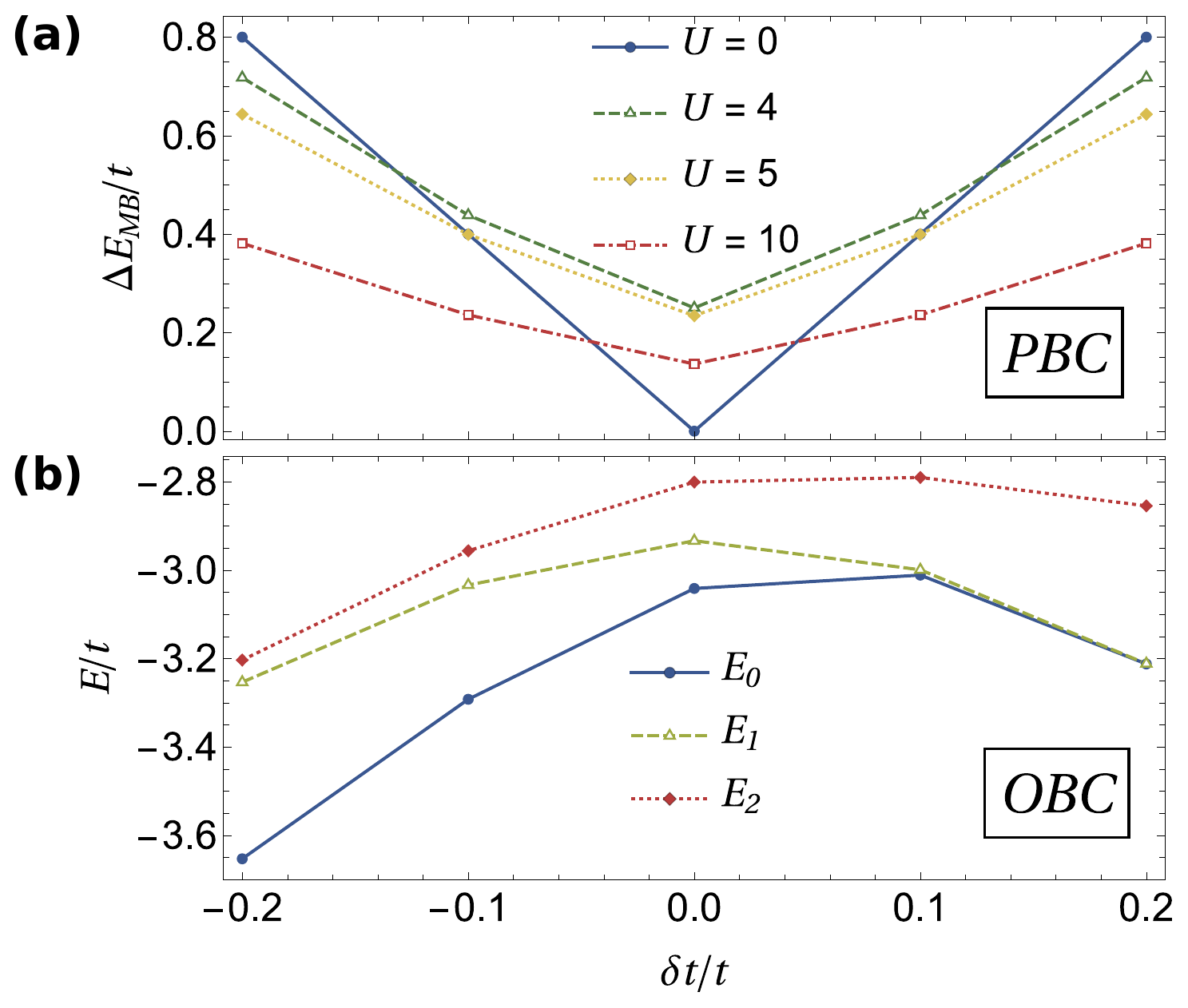}
\caption{Eigenenergies of a chain of $L=12$ sites at zero magnetization $(S_z=0)$: (a) Many-body gap for PBC in dependence of dimerization $\delta t/t$. The gap shows a minimum at $\delta t=0$ independent of $U$ indicating a topological phase transition. (b) Energies of the first three eigenstates for OBC in the strong coupling limit $U=10$. The ground state degeneracy changes upon variation of the dimerization.}
\label{spec_dt}
\end{center}
\end{figure}

We begin by studying the effect of repulsive interactions $U>0$ on both addition and many-body spectrum of an SSH-chain in the topological phase at $\delta t/t=0.5$. Both spectra are shown in Fig.\,\ref{spec_U} for different boundary conditions. 
As for a linear chain ($\delta t=0$) the SSHH model features a charge gap proportional to $U/t$, called the Mott gap, which is clearly visible in the addition spectrum in Fig.\,\ref{spec_U}(a) and explicitly shown in the inset. Since the energetic cost of charge excitations increases with $U/t$, the charge gap shows no local minimum. 
In contrast to the regular Hubbard model which exhibits gapless spin-excitations in the bulk, the spin-sector of the dimerized chain is gapped in the presence of onsite interactions. As shown in the inset of Fig.\,\ref{spec_U}(b), the spin gap of the SSHH model remains finite but decreases monotonically as $\sim 4t/U^2$.
The absence of a local minimum in the bulk gap of both charge and spin sectors indicates absence of an interaction-driven topological phase transition in the SSHH model.

The robustness of the topological SSH-phase against repulsive onsite interactions is further supported by the finite ground state-degeneracy of the open chain due to zero-energy edge states. In Fig.\,\ref{spec_U}(b) we show the energies of the first four eigenstates with zero magnetization for OBC and $\delta t/t=0.5$. 
In the limit of long chains the ground state of the SSH model ($U=0$) is four-fold degenerate corresponding to the four possibilities of distributing two spin-1/2 of opposite sign among two edge sites. In finite systems this degeneracy is slightly lifted once both edge states hybridize to form bonding and anti-bonding states (see Fig.\,\ref{fig:finite_size_spectrum}(d)). Since this energy-splitting is a finite-size effect and is known to vanish in the thermodynamic limit, we can treat these states as degenerate for the purpose of identifying the presence of topological zero-energy excitations. In fact, the long-chain behaviour can be achieved in finite systems by increasing the dimerization as to sufficiently suppress hybridization and localize each edge states on one end of the chain. 
Since two of these four states are comprised of one empty and one doubly occupied edge, their energy increases for $U>0$, leaving two degenerate ground states 
each with singly occupied edges, respectively. 
The finite degeneracy indicates the existence of two zero-energy states, which should be considered as spin excitations, signalling an interacting topological phase at $\delta t>0$.

\begin{figure}[t!]
\begin{center}
    \centering
    \includegraphics[width=\linewidth]{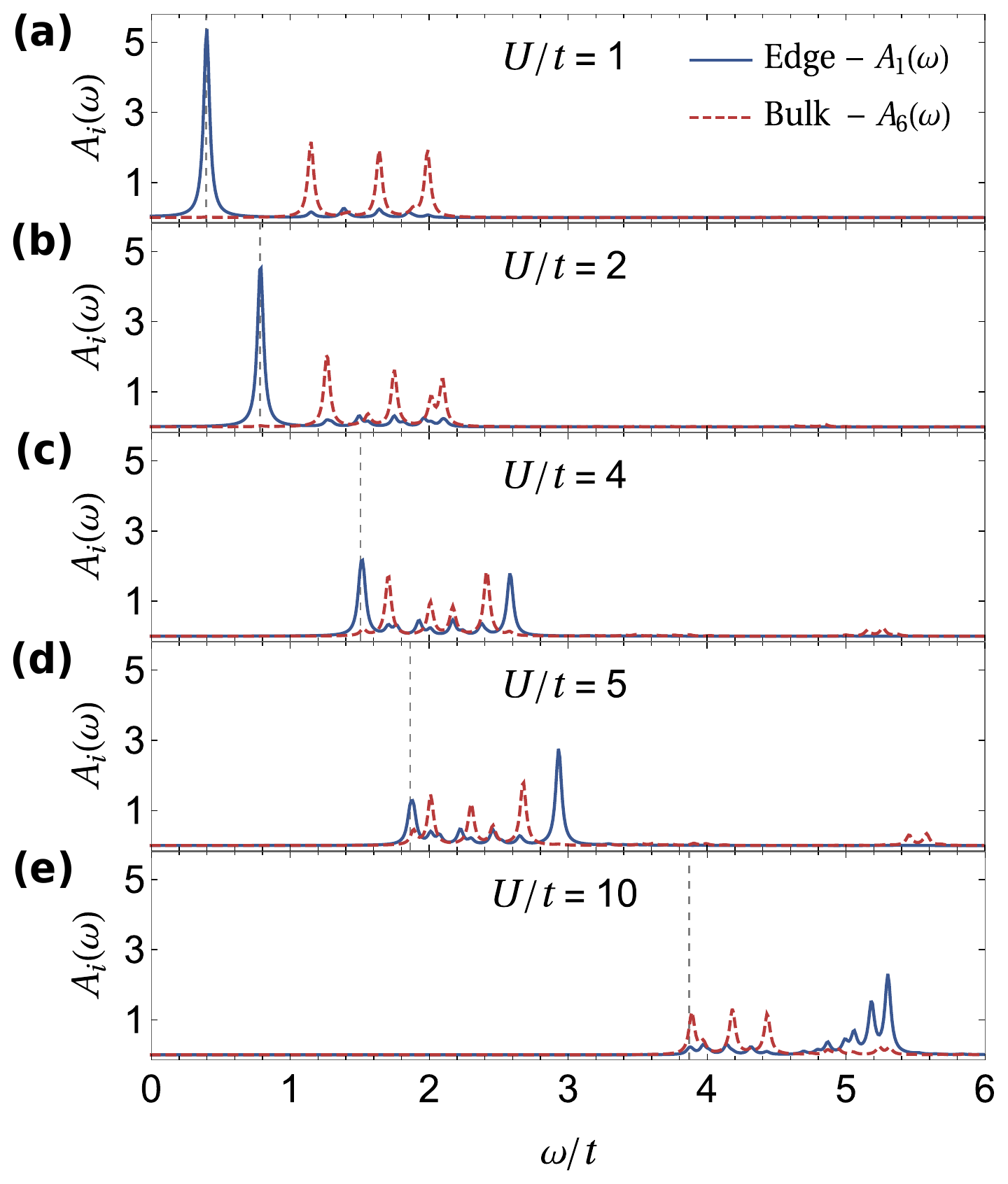}
    \caption{SPSF of edge- (blue, solid) and bulk (red, dashed) sites for (a) $U/t=1$ (b) $U/t=2$ (c) $U/t=4$ (d) $U/t=5$ (e) $U/t=10$. The calculations are done for a chain of $L=12$ sites in the topological phase ($\delta t/t=0.5$). Only positive energies are shown since particle-hole-symmetry implies $A_i(-\omega) = A_i(\omega)$. The gray dashed lines indicates the addition energy $E_\text{add}(N+1)$ corresponding to the ground state transition $|\Psi_0^{N}\rangle \to |\Psi_0^{N+1}\rangle$.}
\label{sfu}
\end{center}
\end{figure}
On the other hand, upon variation of the dimerization instead of the interaction the many-body gap with PBC in Fig.\,\ref{spec_dt}(a) shows a clear local minimum at $\delta t=0$ for all $U/t\leq 10$. Furthermore, a reduction of $\delta t/t$, analogous to the SSH model, lifts the ground state degeneracy and results in a unique ground state as expected for a trivial insulator (cf. Fig.\,\ref{spec_dt}(b)). We note that while the standard Hubbard model $(\delta t=0)$ is gapless in the thermodynamic limit, the gap closing point is shifted for finite system sizes and the degeneracy is already lifted for $\delta t>0$. Both findings strongly indicate a topological phase transition upon varying $\delta t/t$ between two gapped interacting phases. 

As shown in Ref.\,\cite{Manmana2012}, the nature of the interacting zero-energy edge states is distinct from those of the SSH model. While the latter hosts single-particle edge states, which are evident from the zero-energy peaks of the spectral function in Fig.\,\ref{fig:finite_size_sf}, the interacting system features collective spin-excitations on its boundary.

The absence of topological single-particle edge states is clearly observable in the SPSF, which is shown in Fig.\,\ref{sfu}, for a topological chain of 12 sites at various interaction strengths and OBC. Due to the particle-hole symmetry at half filling, it is sufficient to show the SPSF for positive energies. The corresponding emission spectrum is given by $A_i(-\omega) = A_i(\omega)$. The gray dashed line indicates the excitation energy corresponding to the ground state transition $|\Psi_0^{N}\rangle \to |\Psi_0^{N+1}\rangle$. Already for weak interaction of $U/t=1$ [Fig.\,\ref{sfu}(a)], the charge excitation on the edge, depicted in blue, is shifted away from zero-energy. As the edge-sites are on average singly-occupied, the addition of another electron leads to a $U$-dependent energy cost
which results in gapped boundary states and the loss of their topological protection in the charge sector. 

Furthermore, interaction-induced side-peaks emerge at larger energies, indicating transition processes into excited eigenstates. Since the total spectral weight is conserved, the emergence of new peaks goes along with a reduction in amplitude of the dominant in-gap excitation.   
\begin{figure}[t!]
\begin{center}
    \centering
    \includegraphics[width=\linewidth]{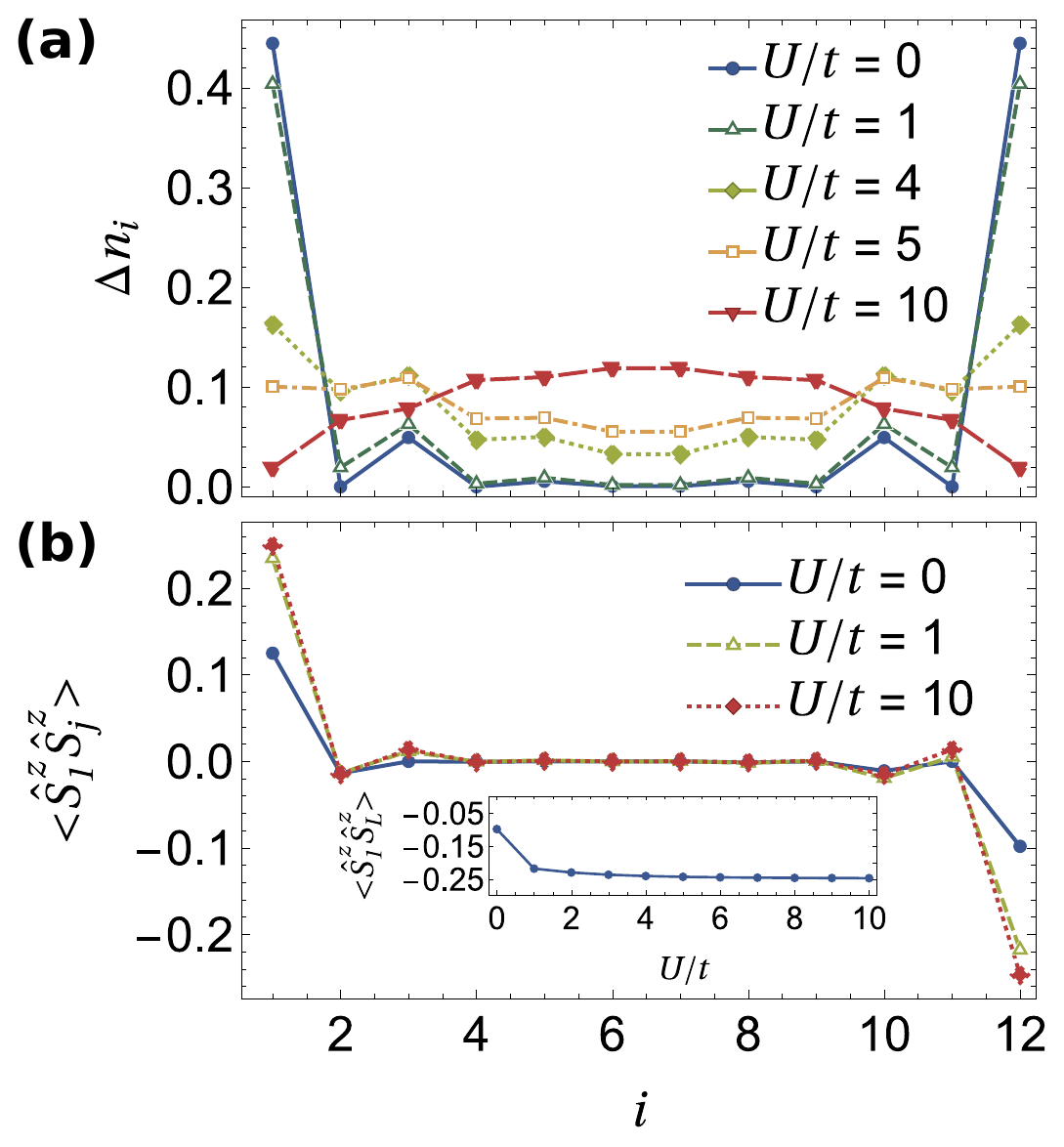}
     \caption{Correlation functions calculated for a 12-site chain with dimerization $\delta t/t = 0.5$: (a) Average probability $\Delta n_i$ to find the inserted electron on site $i$. (b) Absolute value of spin-spin correlation between site $1$ and $i$. Inset: Spin-spin correlation between edge sites in dependence of $U/t$. }
\label{cf}
\end{center}
\end{figure}
Despite the transfer of spectral weight, at weak to moderate interactions the edge excitation is distinctly quasiparticle-like and its spectral weight remains mostly localized at the excitation energy of the ground state transition [Fig.\,\ref{sfu}(a)-(c)].

Since the energy shift due to the repulsive interactions is experienced by both edge and bulk states (red dashed lines),
for sufficiently small values of $U$ the edge states remain within the bulk gap which strongly suppresses hybridization between the edges and the rest of the chain. Interestingly, the interaction-induced energy shift differs in size for edge and bulk, being stronger at the boundary. While electron addition at all sites inflicts double occupation, the kinetic energy is maximized in the center of the chain which
counteracts the widening of the bulk gap.

At a critical interaction strength $U_c/t\approx 5$ [Fig.\,\ref{sfu}(d)] the energy of the ground state transition (gray dashed line) sufficiently overlaps with those of the bulk bands which lifts the in-gap protection and opens a decay channel between edge and bulk. Indeed, Fig.\,\ref{spec_U}(a) shows that the addition energies for OBC and PBC converge at $U_c/t$. At this state the dominant edge excitation is no longer between the ground states and its energy exceeds those of the bulk. We note that the specific value $U_c/t$ is dependent on the chain length $L$.
In the strong coupling limit [Fig.\,\ref{sfu}(e)] the spectral weight of the edge excitations is spread over a range of eigenstates which indicates the eventual breakdown of the quasiparticle description.

Another clear indication of this transition is given by inspection of the probability to find an added electron on site $i$, given by $\Delta n_i(N)=\langle \Psi_0^{N+1}| \hat{n}_i |\Psi_0^{N+1}\rangle - \langle \Psi_0^{N}| \hat{n}_i |\Psi_0^{N}\rangle $\,\cite{Guo2011,Barbiero2018c}. 
In Fig.\,\ref{cf}(a) one can see that at $U=0$ the added electron mainly occupies the zero-energy edge states. Edge-population remains dominant in the previously identified quasiparticle regime $U<U_c\approx 5t$. Only at $U\geq U_c$, the occupation of bulk sites becomes more likely. In the strong coupling limit $U/t=10$, the probability is maximized in the center of the chain. 

The nature of the interacting zero-energy states on the boundary is revealed by the spin-spin correlation function $\langle \hat{S}^z_1 \hat{S}^z_i \rangle $ where $\hat{S}^z_i= \frac{1}{2}\left(\hat{n}_{i,\uparrow}-\hat{n}_{i,\downarrow}\right)$ is the local magnetization on site $i$. It maps out the correlations between unpaired spins on different lattice sites of the half-filled ground state. As shown in Fig.\,\ref{cf}(b) at $U=0$ (blue curve) the correlation between an edge and a bulk spin is strongly suppressed as expected for localized boundary states. The magnitude of the correlation function on the edge is on the order of $0.125$ as two of the four edge configurations involve doubly occupied sites and hence do not contribute to $\langle \hat{S}^z_1\hat{S}^z_i \rangle $. For finite interactions $U>0$, the reduced ground state degeneracy leads to a strong increase in the correlation between unpaired edge spins and its magnitude approaches the maximum value of $0.25$ in the strong coupling limit. Further the negative sign of the correlation function $\langle \hat{S}^z_1\hat{S}^z_L \rangle$ (inset of Fig.\,\ref{cf}(b)) between both edges shows that the topological edge states of the interacting phase are strongly-correlated pairs of opposite spins, each located on a different edge site. 

\begin{figure}[b!]
\begin{center}
    \centering
         \includegraphics[width=\columnwidth]{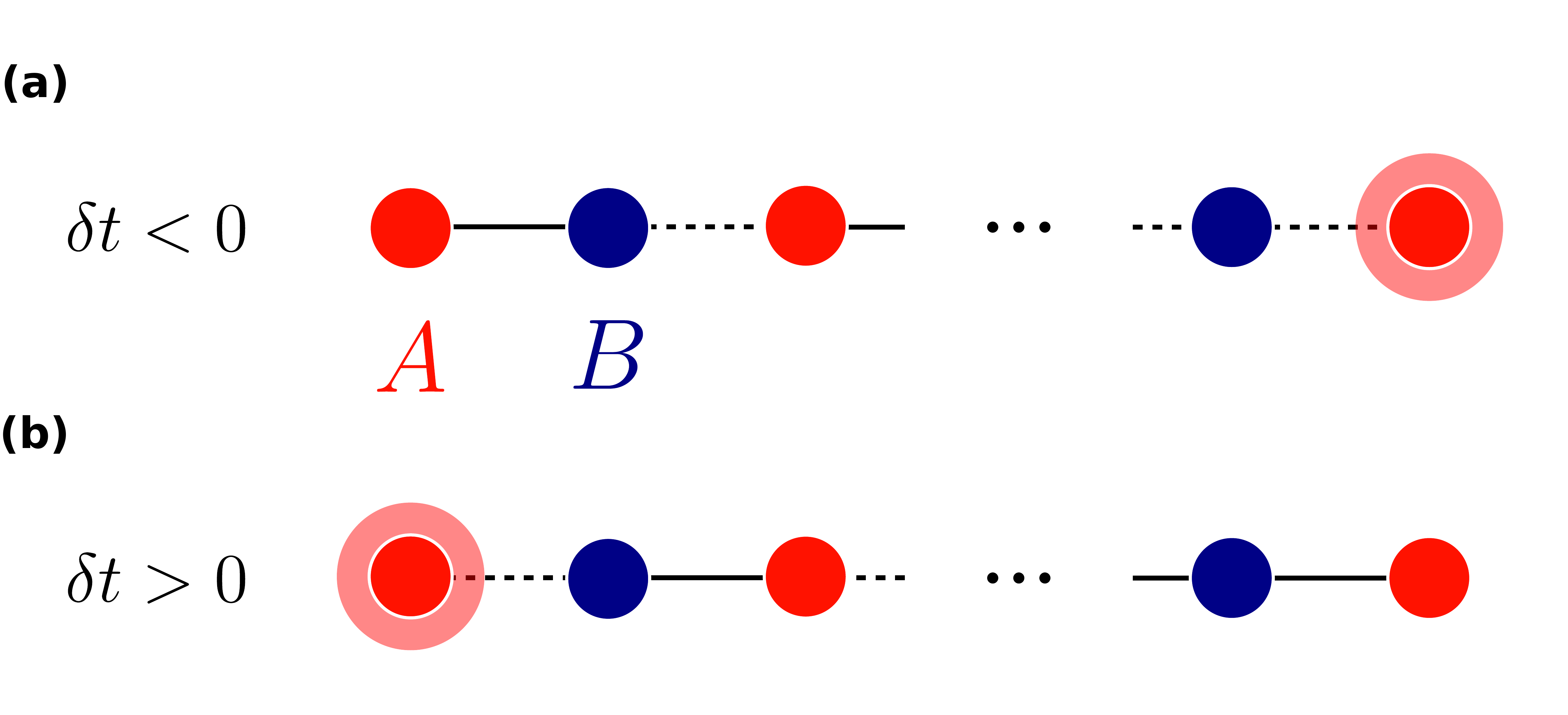}
\caption{
The two different configurations of the SSH model for $L$ being odd. The sublattices $A$ and $B$ are coloured as red and blue sites, respectively. For negative dimerization $\delta t<0$ the zero-energy state is located on the right end of the chain (a) and moves to the left edge once $\delta t$ changes its sign (b). For both configurations the edge state has only support on a single sublattice.}
\label{fig:ssh_odd}
\end{center}
\end{figure}

\section{Odd chains  \label{sec::odd_chains}}
\begin{figure}[t!]
\begin{center}
    \centering
         \includegraphics[width=\columnwidth]{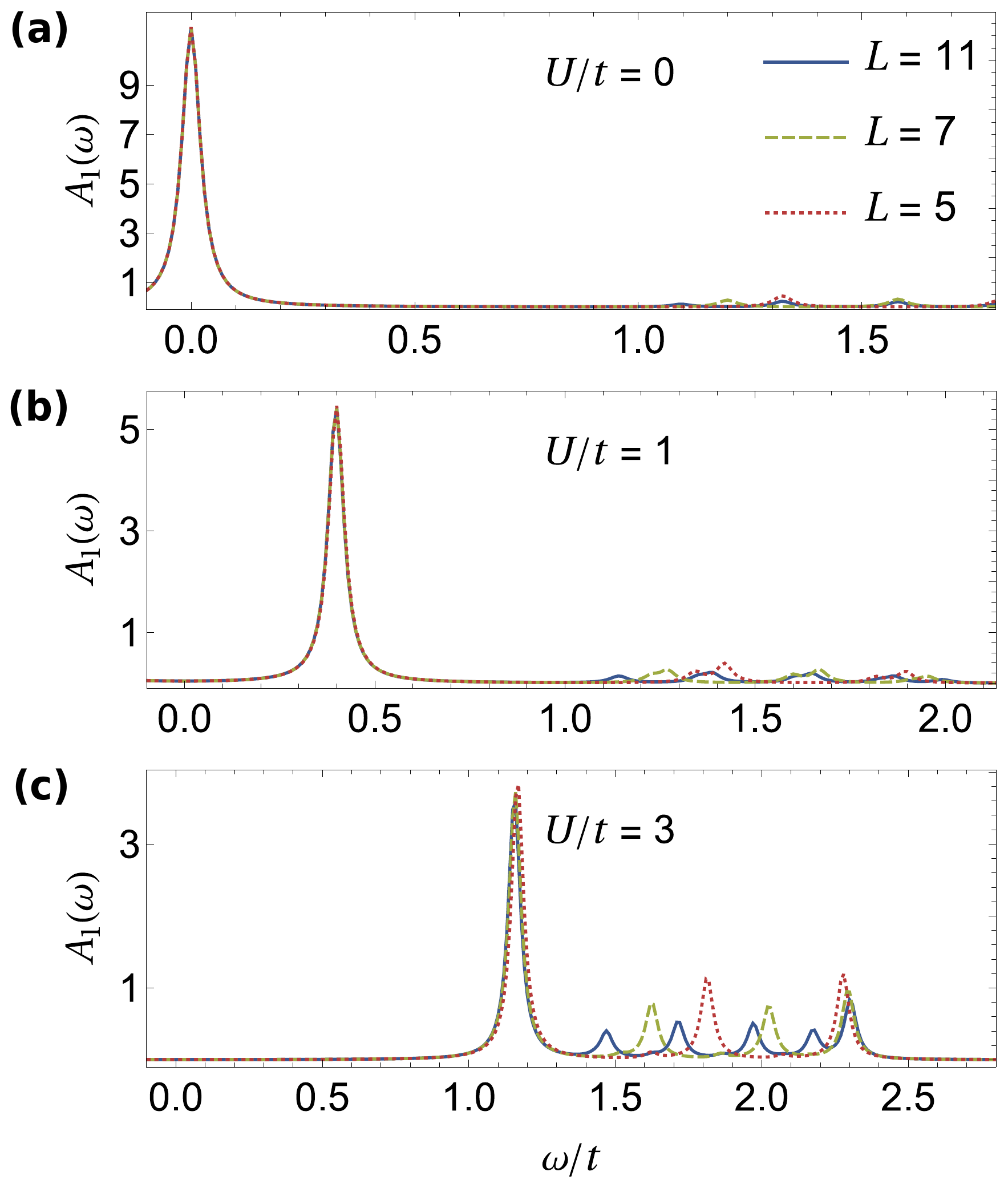}
\caption{Edge single-particle spectral function of for odd chains of different system sizes at dimerization $\delta t/t=0.5$. }
\label{fig:odd_sf}
\end{center}
\end{figure}

The edge physics of the SSH model change considerably when the number of lattice sites is odd instead of even. As schematically shown in Fig.\,\ref{fig:ssh_odd}, for any finite dimerization $|\delta t|>0$ an odd chain features a single zero-energy edge state.
This is a consequence of a theorem first proven by Sutherland \,\cite{Sutherland1986} and Lieb \,\cite{Lieb1989} which states that for a bipartite lattice with sublattices $A$ and $B$ of different sizes, i.e., $N_A\neq N_B$, the number of zero-energy states is given by $|N_A-N_B|$ and their wave functions have vanishing components on the sublattice with fewer sites. 
That is, in contrast to even chains, the edge state is pinned to zero energy even for finite systems and depending on the sign of the dimerization, either located on the left ($\delta t>0$) or on the right ($\delta t<0$) edge of the chain. 
\begin{figure}[t!]
\begin{center}
    \centering
         \includegraphics[width=\columnwidth]{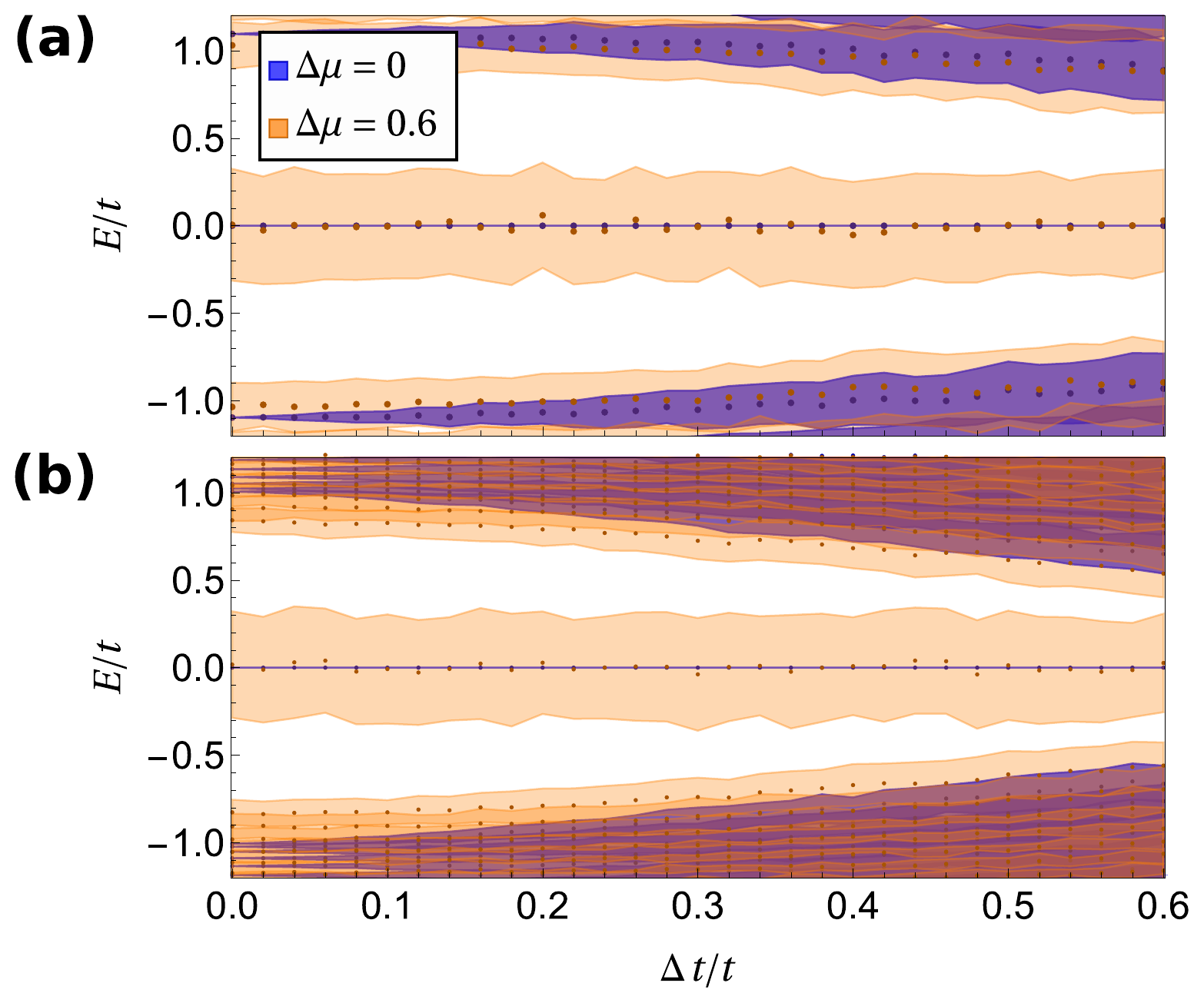}
\caption{Disordered energy spectra averaged over 150 random disorder realizations of a dimerized chain ($\delta t/t=0.5$) of 11 sites (a) and 49 sites (b), respectively. The standard deviation of the disorder averaging is depicted by blueish ($\Delta \mu=0$) and yellowish ($\Delta \mu=0.6$) shaded areas. The edge state for $\Delta \mu=0$ ($\Delta \mu\neq0$) is pinned to $E=0$ (scattered away from $E=0$).}
\label{fig:odd_disorder}
\end{center}
\end{figure}
As shown in Fig.\,\ref{fig:odd_disorder}, the sublattice-polarization of its wave function makes the edge state less susceptible to local disorder as its energy is only affected by random onsite potentials confined to a single sublattice while the hybridization in even chains exposes the edge states to disorder on both sublattices. 
This increased robustness makes odd SSHH chains auspicious candidate systems for the experimental study of topological edge physics on artificial dopant lattices.
Furthermore, due to the lack of hybridization the edge state of the odd chain is substantially less sensitive to the system size as shown by the edge SPSF in Fig.\,\ref{fig:odd_sf} for different chain lengths. The ground state excitation peak barely varies with the number of lattice sites as long as it carries the dominant part of the spectral weight. This further emphasizes the utility of odd chains as they feature clear topological edge excitations at experimentally feasible chain lengths as small as $L=5$. Only once interaction-induced side peaks at higher eigenenergies emerge the size-dependent level-spacings become apparent [Fig.\,\ref{fig:odd_sf}(c)].

%%%%%%%%%%%%%%%%%%%%%%%%%%%%%%%%%%%%%%%%%%%%%%%%%%%%%%%%%%%%%%
%
%                   D I S C U S S I O N
%
%%%%%%%%%%%%%%%%%%%%%%%%%%%%%%%%%%%%%%%%%%%%%%%%%%%%%%%%%%%%%%
\section{Discussion  \label{sec::discussion}}

\subsection{Dopant-lattice experiments}

Electrons bound to phosphorus dopant atoms in silicon form an excellent candidate to emulate the Fermi-Hubbard model. They can be placed with atomic precision using scanning tunneling lithography to form arrays of quantum dots embedded in fully epitaxial silicon\,\cite{Fuechsle2012b}. Each quantum dot consists of a cluster of a few dopants and can represent a site of the chain, and a wide range of ratios between the tunnel coupling, charging energy (which is related to the Hubbard interaction in our model) and temperature can be engineered and explored. The distance between the sites mainly governs the tunnel coupling $t$, and placing sites at a few nanometers from each other results in very large tunnel couplings above 10\,meV\,\cite{He2019}, well above the thermal energy given that dopants can easily be cooled below 100\,mK (equivalent to 8.6\,$\mu$eV). This situation is therefore very favourable to access the regime of low effective temperatures to preserve correlations and quantum fluctuations\,\cite{Salfi2016,Dusko2018c}, which is challenging to achieve in other quantum simulation platforms\,\cite{Georgescu2014}.

The ability to control the hopping amplitude through variation of the inter-dopant distance allows to tune both the dimerization $\delta t$ as well as the Hubbard interaction $U$ through the ratio $t/U$, making the SSHH Hamiltonian a promising model for quantum simulation on artificial dopant lattices. In addition the number of dopants in each cluster can be used to tune the charging energy from a few to hundreds of meV\,\cite{Weber2014a,Fuechsle2010}.
As a result, dopants in silicon can achieve the threshold $U_c\sim5t$ mentioned above below which signatures of coherent edge states are preserved for the system size considered in this work. Also, disorder in tunnel coupling can be mitigated to avoid mixing between bulk and edge states (see Fig.\ref{fig:ssh_disorder}) using preferential crystallographic placement\,\cite{Voisin2020}. We note that dopants in silicon also present a large inter-site Coulomb repulsion energy $V$ on the order of $t$\,\cite{Kiczynski2022}. Here, we have not taken this term into account following recent work that has predicted a minimal influence on the half-filling picture~\cite{Le2017}, and leave the detailed analysis for future work.

Recently, transport in dimerized chains of 10 dopant-based quantum dots in silicon tuned at quarter-filling was experimentally achieved to evidence a pattern of conductance peaks consistent with the presence of edge states\,\cite{Kiczynski2022}. This breakthrough experiment demonstrates the relevance of dopants in silicon to emulate Fermi-Hubbard problems, with future experiments to focus on exploring different filling factors, spin physics in finite magnetic fields, dynamics and scaling up beyond classical computational capabilities. Scanning tunnelling microscopy is another experimental avenue to probe atomic systems in the solid-state, enabling a direct visualization of edge states\,\cite{Nadj-Perge2014,Drost2017,Kempkes2019}, a capability that is missing in transport experiment. Spatially resolved spectroscopy of single\,\cite{Salfi2014a,Voisin2015} and coupled dopants\,\cite{Salfi2016,Voisin2020} in silicon has been achieved in the single electron transport regime, with the ability to pinpoint the dopant's exact lattice site position\,\cite{Usman2016}. Recent progress takes the direction of integrating this quantum state imaging technique within a device architecture including local gate electrodes\,\cite{Ng2020}. The ability to engineer and locally probe dopant-based arrays in silicon motivates our study focusing on identifying the signatures of an SSH chain that can be expected in this experimental framework in the presence of finite interactions.

\begin{figure}[t!]
\begin{center}
    \centering
         \includegraphics[width=\columnwidth]{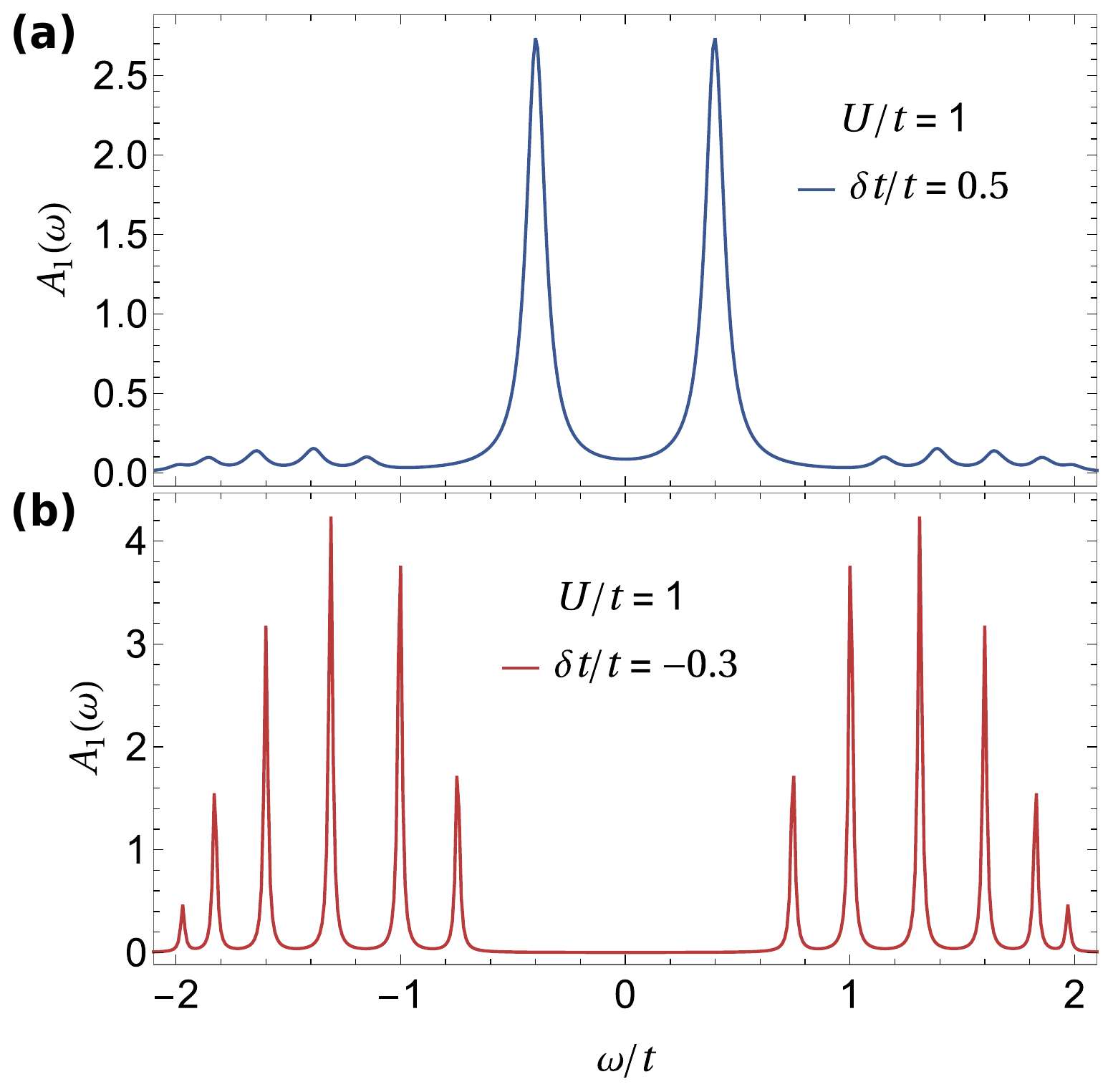}
\caption{
Example of what could be observed in a scanning tunneling experiment for a dopant lattice, measured at the end site of the chain. (a) Topological regime for $\delta t/t=0.5$, $L=12$ and $U=1$. (b) Trivial regime for $\delta t/t=-0.3$, $L=12$ and $U=1$.}
\label{fig:experiment}
\end{center}
\end{figure}

At half filling, a typical local $dI/dV$ measurement for an edge site will reveal the edge states as gapped double peaks, when the system is topologically non-trivial ($\delta t>0$). The distance between the peaks is roughly $U/t$. This is in stark contrast to the trivial regime where the spectral properties of an edge site are essentially indistinguishable from bulk sites. Two characteristic examples--one for the topological and one for the trivial regime--are shown in Fig.\,\ref{fig:experiment} with parameters $L=12$, $U/t=1$, $\delta t/t=0.5$, $\eta=0.05$ (a) and $L=12$, $U/t=1$, $\delta t/t=-0.3$, $\eta=0.01$ (b). 
The split edge-peaks for finite $U$ may become comparable to the peaks stemming from the bulk in the trivial case; in the topological case other bulk peaks are, however, strongly suppressed [Fig.\,\ref{fig:experiment}\,(a)] in contrast to the trivial case [see Fig.\,\ref{fig:experiment}\,(b)]. As long as $U/t < 5$ scanning tunneling measurements should be able to distinguish trivial and topological regimes for the considered chain length.
Chains with an odd number of sites benefit from the absence of hybridization effects irrespective   of chain length. Nonetheless, the edge peak will split for any finite $U/t$.

Higher-order topological phases\,\cite{Benalcazar2017} provide another avenue for dopant lattices to realize exotic states of matter, essentially as two-dimensional extensions of the SSH or SSHH models. Corner states will be observable on appropriately chosen lattice geometries\,\cite{Bunney2022}.

\subsection{Bulk obstructed phase}
Originally introduced to describe solitons in polyacetylene, the SSH model was later found to be a prototype for a 1D TI\,\cite{Ryu,Shen2012,Kane2013,Wang2015f,Asboth2016a}. Recently it was argued that the term {\it bulk obstructed topological insulator} might be more appropriate\,\cite{Bradlyn2017a,Khalaf2021}. While for a traditional TI the non-trivial phase can be identified by the absence of an atomic limit, electronic phases in one spatial dimension ($d=1$) can always be represented in terms of localized Wannier states. Hence in the SSH model the topological distinction between both phases is more subtle and can be found in the relative position between Wannier centers and the average ion location. Inversion symmetry forces the Wannier orbitals, which in the SSH model reside on the strong bonds, to be localized at one of two inversion centers which is either in the center or at the edge of the unit cell. The occupation of the latter characterizes the non-trivial phase and constitutes an obstructed atomic limit which cannot be connected to the trivial limit without either a bulk gap closing at $\delta t = 0$ or the explicit breaking of inversion symmetry. Note that the identification of the non-trivial phase is only unique up to the choice of unit cell. Since a change in the sign of the dimerization amounts to a global shift of the strong bonds by half a lattice vector, moving the origin of the unit cell by the same amount exchanges the atomic positions of the trivial and obstructed atomic limits and therefore the role of the trivial and topological phase. Nevertheless, independent of the reference frame both phases are separated by a true topological phase transition at the gap closing point which is characterized by a quantized topological invariant, the winding number of the Bloch Hamiltonian in the Brillouin zone, and the emergence of symmetry-protected zero-energy states on the system boundary. 
Since the existence of both edge states and bulk gap closing are not affected by the presence of an (obstructed) atomic limit, it seems well justified in the present context to consider the SSH model a TI. Clearly the above discussion is mostly of relevance to experts in the field of topological states of matter; it does not represent a roadblock for the attempt to experimentally realize the SSHH model as a prototype of an interacting, topological model.

\subsection{Correlated topological insulators}

As just explained, 1D TIs such as the SSH model are bulk obstructed models. Ignoring this for a moment, we can turn to the question how the SSHH model compares to other correlated or interacting TI systems. Most importantly, 1D fermonic phases have been fully characterized. The $\mathbb{Z}$ classification of the BDI class changes to $\mathbb{Z}_4$ in the presence of interactions (note that we exclude superconductivity here, otherwise it was $\mathbb{Z}_8$)\,\cite{Fidkowski2010c}.
Such rigorous results are not available in higher dimension $d>1$. Nevertheless, for many paradigmatic models interacting phase diagrams were derived by means of analytical and numerical methods\,\cite{Rachel2010a,Wu2012}.  Finite-size effects in correlated TI's were investigated as well\,\cite{Tada2012}. Conventional ordering tendencies are ubiquitous in $d=2,3$. For instance, TI models on bipartite lattices tend to order antiferromagnetically in the presence of non-negligible interactions. Correlated TI models which break the $U(1)$ spin symmetry are likely to also induce more exotic magnetic orders such as spin spirals. Last but not least the interplay of non-trivial band-topology and electron-electron interactions was claimed to stabilize topologically ordered, spin-liquid type phases\,\cite{Rachel2018a}.

\subsection{Topological phase diagram and Haldane phase}

In the literature, several works\,\cite{Hida1992,Manmana2012,Yoshida2014,Wang2015f,Ye2016,Sbierski2018c,Barbiero2018c,Le2020} comment on the phase diagram of the SSHH model or even sketch it explicitly\,\cite{Manmana2012,Wang2015f,Ye2016}. Moreover, there are several works where additional Heisenberg terms were included to the SSHH Hamiltonian, in order to demonstrate the connection to well-known spin chain results.
Nevertheless, we feel that the reader may benefit from a summary of all these results and from a discussion how they fit together. The main conclusion will be that the topological phase ($\delta t>0$) of the SSHH model is adiabatically connected to the Haldane phase of spin-1 chains\,\cite{Haldane1983}.

As we have seen in section\,\ref{sec::interacting}, single-particle edge states of the SSH model (i.e., SSHH model at $U=0$) change immediately when finite interaction $U>0$ are included and split from $E=0$ into two peaks at some finite $\pm E$. Also, single-particle edge states at $E=0$ become collective $E=0$ edge states, reminiscent of spin-excitations. This change coincides with a change of the degeneracy of the entanglement spectrum\,\cite{Yoshida2014,Ye2016}: at $U=0$ the degeneracy for 
PBCs is 16-fold, and only 4-fold at any non-zero $U$ (here all magnetization sectors were considered). A similar result is found by computing the von Neumann entanglement entropy\,\cite{Wang2015f}. In contrast, the winding number $N_1$ of Ref.\,\cite{Manmana2012} does not change its finite value when tuning from $U=0$ to $U>0$.

Next let us consider the limit of strong coupling. A two-site Hubbard-model $\sum_\sigma \pm t( \hat{c}_{1\sigma}^\dag \hat{c}_{2\sigma}^{\phantom{\dag}} + {\rm H.c.}) + U(\hat{n}_1 + \hat{n}_2)$ in the limit $U\gg t$ corresponds in second order perturbation theory to the Heisenberg term $J \vec S_1\cdot\vec S_2$ with $J=4(\pm t)^2/U$. Thus the SSHH model in this limit corresponds to the dimerized Heisenberg model\,\cite{Hida1992,Wang2013a}
\begin{equation}\label{dimerizedHM}
H= \sum_{i=1}^{L/2} \big[ J \vec S_{2i-1}\cdot\vec S_{2i} + J'  \vec S_{2i}\cdot\vec S_{2i+1} \big]
\end{equation}
with
\begin{equation}
    J^{(')} = \frac{4(t\mp\delta t)^2}{U}\ .
\end{equation}
For the limit of decoupled dimers in the trivial phase, $\delta t=-t$, also the spin model\,\eqref{dimerizedHM} describes decoupled singlet bonds, $J\not=0$ and $J'=0$. In contrast, for the decoupled-dimer limit in the topologically non-trivial regime, $\delta t=t$, the spin model\,\eqref{dimerizedHM} features again a ground state consisting of decoupled singlet bonds; however, due to $J=0$ and $J'\not= 0$ the spin-1/2's at the chain ends are not coupled into singlets and remain as {\it dangling spins}, a situation well known from the spin-1 Heisenberg or Affleck-Kennedy-Lieb-Tassaki (AKLT)\,\cite{Affleck1987,Affleck1988} chains.

The relationship to the spin-1 chain becomes even more obvious when we allow $J$ to become negative\,\cite{Manmana2012} (which is obviously not the strong-coupling limit of the SSHH model anymore) while $J'>0$. In the limit $J\to-\infty$, neighboring spin-1/2's combine symmetrically and form spin-1 degrees of freedom. That is, a dimerized spin-1/2 Heisenberg chain with $L$ sites become an isotropic spin-1 Heisenberg chain with $L/2$ sites. The spin-1 Heisenberg chain features the ''Haldane gap''\,\cite{Haldane1983} and constitutes the prototype of a symmetry-protected topological (SPT) phase\,\cite{Pollmann2010a}. In Ref.\,\cite{Manmana2012}, Manmana \textit{et al.} showed that there is no phase transition when tuning from $J\sim -\infty$ to $J=0$ to $0<J<J'$, i.e., the many-body gap remains finite.
At $0<J=J'$ one reaches the isotropic spin-1/2 Heisenberg chain, where the many-body (spin) gap closes due to gapless spinon excitations.

The relation between the dimerized Heisenberg chain with $J<0$ and with $J>0$ and its connection with the Haldane phase was already shown by Hida in his 1992-paper\,\cite{Hida1992}, long before the notion of SPT phases was born. He demonstrated that the string-order parameter characteristic for the Haldane phase is finite for positive and negative $J$ as long as $J'>0$ and $J<J'$, regardless of the sign of $J$. He further computed the energy gap which remained finite for $-\infty < J < J'$. Most convincingly, for $J'\to -\infty$ the literature values of both string order parameter and many-body gap of the isotropic spin-1 Heisenberg chain were reached. 

In summary, the trivial phases of the SSH and SSHH models are connected to the trivial phase of the dimerized Heisenberg chain. In the topological regime, both the non-interacting SSH and the interacting SSHH models are topologically non-trivial; however, the $U=0$ line is different from the $U>0$ phase as signalled by different degeneracies in the entanglement spectrum and the nature of the zero-energy edge states. The topological phase of the SSHH model is adiabatically connected to the spin-1 chain and constitutes, hence, another example of a Haldane-gap model.

%%%%%%%%%%%%%%%%%%%%%%%%%%%%%%%%%%%%%%%%%%%%%%%%%%%%%%%%%%%%%%
%
%                   C O N C L U S I O N 
%
%%%%%%%%%%%%%%%%%%%%%%%%%%%%%%%%%%%%%%%%%%%%%%%%%%%%%%%%%%%%%%

\section{Conclusion  \label{sec::conclusion}}
In this work we studied the effect of repulsive onsite interactions on the topological edge states of the spinful SSH model by means of ED. 
Motivated by recent progress in the fabrication of dopant-based quantum simulators we focused on topological signatures which are experimentally detectable. 
To this end we present the real space resolved SPSF as a tool to reveal non-trivial topology in correlated systems. As the generalization of the LDOS to interacting systems the SPSF is directly accessible to local measurement techniques like scanning tunneling methods. 
By comparison with large chains in the free-fermion limit we confirmed the persistence of topological edge excitations observed within the SPSFs of the system sizes accessible to ED. Further, we showed the robustness of these edge states in the presence of random bond and onsite disorder.
Our analysis of the interacting model showed that repulsive Hubbard interactions gap out the charge sector and the single-particle edge states are no longer pinned to zero energy. Instead the topological states involve two correlated edge spins of opposite direction. 
Nevertheless, we identify a range of interactions for which quasiparticle excitations at the boundary, remnants of the topological SSH states, are protected by a bulk gap and as such serve as clear indicators for the topological nature of the underlying ground state. 
We emphasise the relevance of this result for the detection of interacting topological phases in single-particle based measurements like STS.
Beyond the interaction threshold of $U_c/t\approx 5 $, where edge and bulk excitation energies align for the here considered chain lengths, we observe the decay of the edge states due to increased spectral weight transfer and a breakdown of the quasiparticle description.
We discussed the topological phase diagram of the SSHH model and its connection to the Haldane phase of spin-1 chains.

\vspace{20pt}
\section{Acknowledgements}
We acknowledge support from the ARC Centre of Excellence for Quantum Computation and Communication Technology (CE170100012), an ARC Discovery Project (DP180102620) and Silicon Quantum Computing Pty Ltd., and an ARC Future Fellowship (FT180100211).

\appendix

%%%%%%%%%%%%%%%%%%%%%%%%%%%%%%%%%%%%%%%%%%%%%%%%%%%%%%%%%%%%%%
%
%                   A P P E N D I X
%
%%%%%%%%%%%%%%%%%%%%%%%%%%%%%%%%%%%%%%%%%%%%%%%%%%%%%%%%%%%%%%
    
\section{Exact Diagonalization\label{app:ED}}
Strongly-correlated materials are among the most intriguing systems in condensed matter physics, displaying exotic phases ranging from high-$T_c$ SC to Mott insulators and quantum magnetism. The theoretical investigation of these many-body phenomena usually starts with a variant of the Fermi-Hubbard model of correlated electrons on a lattice. Notwithstanding the great interest in these models, analytical solutions and controlled numerical methods are rare and often limited to specific cases.

ED, a numerical method for solving the many-body Schroedinger equation which involves no approximation and produces the exact eigenenergies and eigenstates of the full interacting Hamiltonian, constitutes an exemption. As such it allows to study static and dynamic correlation functions of the ground state and, in principle, for finite temperatures. It further serves to benchmark the predictions of other methods. The drawback of this approach is that computational time and memory requirements are proportional to the dimension of the Hilbert space $d$. For an interacting many-body system it grows exponentially with the system size, i.e.,  $d = d_l^{L} $ where $d_l$ is the dimension of the local Hilbert space at a given site and $L$ is the total number of lattice sites. In the case of the spin-1/2 Hubbard model, there are four possible states per lattice site given by $|\emptyset\rangle,|\uparrow\rangle,|\downarrow\rangle$ and $|\uparrow\downarrow\rangle$ leading to a total dimension of $d=4^{L}$. 

One way to reduce the size of the Hilbert space is to exploit unitary symmetries of the system for which the Hamiltonian matrix is block diagonal with each block being labeled by the associated conserved quantum number.  
E.g. in the case of $U(1)$ charge symmetry, the total particle number $N$ is conserved and the Hamiltonian decomposes into independent particle sectors, each of which can be diagonalized separately. To do so, the first step is to express the relevant block of the Hamiltonian matrix in a basis of Slater determinants $|\Phi_I\rangle$. 
By use of the occupation number representation, i.e., 
\begin{align}
|\Phi_I\rangle = |n_L, n_{L-1},...,n_1\rangle_I = \prod_{i=1}^{L} \left(\hat{c}^\dagger_i  \right)^{n_i}|0 \rangle,     \label{sd}
\end{align}
with site-occupations $n_i\in \{0,1\}$, each basis state can be uniquely associated with the binary representation of an integer number $I$\,\cite{Pavarini2019}. The action of a creation (annihilation) operator $\hat{c}^\dagger_i$ ($\hat{c}_i$) on a basis states then simply translates to a flip of the $(i-1)^{\text{th}}$ bit from 0 to 1 (1 to 0) times a potential minus sign stemming from fermionic anti-commutation relations. Using a procedure first outlined by Lin\,\cite{Lin1990}, the above mentioned basis-representation can be extended to spinful electrons in a straightforward manner\,\cite{Lin1993}. By application of the Hamiltonian to each basis state, all matrix elements $H_{II'}= \langle \Phi_I | \hat{H} | \Phi_{I'}\rangle $ are successively generated. 
Once the matrix is complete, a diagonalization routine is applied to obtain the eigenvalues and eigenvectors. 

To this end we employ the Lanczos algorithm\,\cite{Lanczos1950} which is an efficient method to obtain the few lowest (or largest) eigenvalues of a sparse Hermitian matrix.   
The main idea is to transform the $d$-dimensional Hamiltonian matrix $H$ into a tridiagonal matrix
\begin{align}
T=
\begin{pmatrix}
a_1 & b_2 & 0 & \cdots & 0 \\
b_2 & a_2 & \ddots &  &\vdots  \\
0 & \ddots & \ddots & \ddots  & \\
\vdots &  & \ddots & \ddots &b_{d_T}\\
0 & \cdots & &b_{d_T} & a_{d_T}  \label{tmatrix}
\end{pmatrix}
\end{align}
of dimension $d_T\ll d$, the eigenvalues of which approximate the ones of the full matrix to arbitrary precision. This is done by expressing the Hamiltonian in the so called \textit{Krylov space} $\mathcal{K}^{d_T}(|\tilde{\Psi}\rangle)=\text{span}(|\tilde{\Psi}\rangle,\hat{H}|\tilde{\Psi}\rangle,\hat{H}^2|\tilde{\Psi}\rangle, ...,\hat{H}^{d_T}|\tilde{\Psi}\rangle)$ of some arbitrary (normalized) trial vector $|\tilde{\Psi}\rangle$. For a half-filled Hubbard chain of $L=12$ sites, the lowest eigenvalue of $T$ converges to the exact ground state energy after only $d_T<100$ iterations\,\cite{Weisse2008}. The computationally most expensive part of the Lanczos routine is the sparse matrix-vector multiplication used to create the Krylov sub-space. For a matrix of $n_\text{nz}$ non-zero matrix elements, it scales with $\sim \mathcal{O}(n_\text{nz}*d)$ leading to an overall complexity of $\mathcal{O}(n_\text{nz}*d_T*d)$ (compared to $\mathcal{O}(d^3)$ for the full diagonalization of dense matrices). Since we are interested in the SPSF, we have to diagonalize the three Hamiltonian sectors of $N-1$, $N$ and $N+1$ particles, respectively. Fortunately, for the half-filled sector, which constitutes the largest block, only the ground state and its energy are needed (see Eq.\,\eqref{GF0}) which reduces the number of needed iterations even further as extremal eigenvalues converge most rapidly\,\cite{Pavarini2019}.

In concluding this section we want to emphasize that, despite its limitation to comparably small system sizes, ED proves to be a valuable method for the study of interacting topological phases. For example, Varney \textit{et al.} found\,\cite{Varney2011h} that for correlated Chern insulators the notion of a topological phase transition, signalled by the closing of the many-body gap, is well-defined even for small clusters accessible to ED. It was further shown that lattices of a few sites in 1D\,\cite{Guo2011,Le2020} and 2D\,\cite{Varney2010h} can exhibit interacting topological edge states and current-current correlations\,\cite{Capponi2015} associated with a non-zero bulk invariant. 
ED was also employed to study the many-body instabilities of a 3D TI surface state along with the possible appearance of Majorana modes\,\cite{Neupert2015}.
These findings demonstrate the relevance of ED calculations for the theoretical understanding of both correlated TIs as well as their experimental realization in dopant lattices, where small system sizes are common.

\section{Single-particle spectral function \label{app:SPSF}}
Here we introduce the main properties of the SPSF in real space and discuss its relevance for the description of measurements in STS experiments. To establish the relationship between the conductivity and the SPSF we consider the general expression for the tunneling current between single-particle states in the tip $\psi_\mu$ and the sample surface $\psi_\nu$ at a voltage difference $V$\,\cite{Bardeen1961} 
\begin{align}
    I &=\frac{2\pi e}{\hbar} \sum_{\mu,\nu}  f(E_\mu)[1-f(E_\nu+ eV)] \nonumber\\ 
    &\times|M_{\mu\nu}|^2 \delta(E_\mu-E_\nu) \label{current1},
\end{align}
with Fermi-function $f(E)$ and the tunneling matrix element $M_{\mu\nu}$ between the tip and surface states. In their seminal works on the theoretical description of the STM\,\cite{Tersoff1983,Tersoff1985}, Tersoff and Hamann showed that for a spherically symmetric tip the matrix elements are proportional to the surface states at the position of the tip $M_{\mu\nu}\propto \psi_\nu(r_0$). At low temperatures, $f(E)$ approaches a step-function and mainly counts states which lie between the Fermi energies of tip and sample and hence contribute to the tunneling. The expression for the current at finite voltages is then given by the convolution of the tip DOS $\rho_t(E)$ and the LDOS of the sample at the tip-position $\rho_s(r_0,E)$\,\cite{Odashima2017} 
\begin{align}
    I \propto \int^{eV}_{0}&d\omega \,\rho_s(r_0,E_F-eV+\omega) \rho_t(E_F+\omega) \nonumber\\
    &\times T(\omega,V), 
\end{align}
where the transmission coefficient $T(\omega,V)$ describes the energy-dependent part of the matrix elements. 
For a metallic tip and sufficiently small voltages, the energy-dependence of $\rho_t$ and $T$ is considered to be weak\,\cite{Wiesendanger1994} such that the profile of the measured conductivity is dominated by the LDOS of the sample's surface, i.e., 
\begin{align}
    \frac{\partial I}{\partial V} \propto \rho_s(r_0,E_F-eV).  \label{conductivity}
\end{align}
To preserve this relation in the context of strong electron correlations one has to generalize the LDOS for the many-body case, which is done by SPSF in real space, which
provides a direct link between theory and experiment as it represents the excitation spectrum associated with the addition and removal of a single electron. 
In other words, given a system in its $N$-particle ground state, the spectral function $A_\alpha(\omega)$ assigns the probability for the system's energy to decrease (increase) by $\omega$ after annihilation (creation) of an electron in the single-particle state $|\alpha\rangle$. For the purpose of studying excitations in real space, we choose the site basis of the lattice $|\alpha\rangle \equiv |i\rangle$. 
As a probability measure, the spectral weight of the spectral function is conserved which is expressed by the normalization condition 
\begin{align}
    \int_{-\infty}^{\infty} \,d\omega A_{i}(\omega) = 1. \label{norm}
\end{align}
As mentioned in Section\,\ref{sec::methods} the SPSF is proprtional to the imaginary part of the retarded Green's function\,\cite{Odashima2017}
\begin{align}
    A_{i}(\omega)= -\frac{1}{\pi} \text{Im} (G^\text{r}_{i}(\omega))\label{SF}
\end{align}
which in its spectral representation is given by
\begin{align}
    G^\text{r}_{i}(\omega) \!=\! \frac{1}{Z}\sum_{mn} |\langle \Psi_m|\hat{c}_i|\Psi_n\rangle |^2 \frac{e^{-\beta(E_m - \mu N_m)}+e^{-\beta(E_n - \mu N_n)}}{\omega +\mu -(E_n -E_m)+i \eta}, \label{GFt}
\end{align}
where $Z=\sum_{m} e^{-\beta(E_m - \mu N_m)}$ is the quantum statistical partition function, $\mu$ is the chemical potential, $\beta=1/T$ is the inverse temperature. 
The convergence factor $\eta>0$ depicts an infinitesimal shift away from the real axis to avoid the singularities of the Green's function, which occur at the excitation energies $\omega=E_n-E_m-\mu$. The corresponding transition amplitudes are non-zero only between states which differ in filling by a single electron, i.e., $N_n-N_m=\pm1$. 
At zero temperature ($\beta \to \infty$) the Boltzmann-factor $e^{-\beta(E_m - \mu N_m)}$ vanishes for all $m$ except for the ground state at $m=0$ where, by appropriate choice of the chemical potential $\mu=\mu_*$, it becomes unity. For the half-filled Hubbard model one has $\mu_*=0$ and the double sum in Eq.\,\eqref{GFt} reduces to 
\begin{align}
    G^\text{r}_{i} (\omega) = &\sum_{n,\sigma} \frac{|\langle \Psi_n^{N-1}|\hat{c}_{i,\sigma}|\Psi^N_0\rangle|^2}{\omega +\mu_* -(E^{N}_0 -E^{N-1}_n)+i \eta} \nonumber \\[5pt] &+\frac{  |\langle \Psi_n^{N+1}|\hat{c}^\dagger_{i,\sigma}|\Psi^N_0\rangle|^2}{\omega +\mu_* -(E^{N+1}_n -E^{N}_0)+i \eta}, \label{GF0}
\end{align}
where the first (second) term corresponds to the single-particle transition into the $n_\text{th}$-eigenstate with $N-1$ ($N+1$) particles upon removal (addition) of an electron. The Green's function is peaked at the corresponding excitation energy and weighted by the associated transition probability. 
The restriction to three particle sectors renders the numerical evaluation of Eq.\,\eqref{GF0} by means of ED a feasible endeavour. 
Further, since the maximum energy of a single particle excitation is on the order of the interaction strength, i.e.,  $E^{N+1}_{n}-E^N_{0} \approx U$, it is sufficient to evaluate the low-energy end of the full spectrum without loss of relevant information.  
This allows for the utilization of the Lanczos routine , described in Appendix\,\ref{app:ED}, and amounts to a substantial reduction in computational effort.

Alternatively to calculating the Lehmann representation Eq.\,\eqref{GF0} by determining the eigenspectrum of the three particle sectors, the Lanczos method can be used to calculate correlation functions like the SPSF directly\,\cite{Fulde1993,Lin1993}. The basic idea of this approach is to run the Lanczos routine twice. In the first run the groundstate $|\Psi_0\rangle$ and its energy $E_0$ are calculated for the half filled sector. Consecutively the groundstate is then used to span the Krylov space in the second run by choosing the initial basis vector as  
$|\tilde{\Psi}\rangle=\hat{a}|\Psi_0\rangle/\sqrt{\langle \Psi_0 |\hat{a}^\dagger \hat{a} |\Psi_0 \rangle}$ with $\hat{a}=\hat{c}_i$ or $\hat{a}=\hat{c}^\dagger_i$, depending on whether excitations involving the removal or addition of an electron are considered. The entries $a_n$ and $b_n$ of the generated tridiagonal matrix Eq.\,\eqref{tmatrix} can then be used to calculate the SPSF in a continued fraction via 
\begin{align}
A_i(\omega) = -\frac{1}{\pi} \text{Im} \left(\frac{\langle \Psi_0| \hat{c}_i^\dagger \hat{c}_i |\Psi_0 \rangle}{z^-+a_1-\frac{b_2^2}{z^-+a_2 -\frac{b_3^2}{\ddots}}} + \frac{\langle \Psi_0| \hat{c}_i \hat{c}_i^\dagger  |\Psi_0 \rangle}{z^+-a_1-\frac{b_2^2}{z^+-a_2 -\frac{b_3^2}{\ddots}}} \right) \label{contfrac}
\end{align}
with $z^\pm=w+\mu\pm E_0 + i\eta$. 
Due to specific choice of the initial vector the algorithm quickly converges to the first $d_T$ lowest excitations and therefore allows to include a larger number of poles in the calculation of the SPSF. As such the evaluation of Eq.\,\eqref{contfrac} provides an advantage in the strongly interacting regime, where excitations of higher energies become relevant.

In the absence of interactions the eigenenergies, i.e.,  the band structure, are independent of the filling fraction and the system remains in an eigenstate after removal or addition of an electron\,\cite{Altland}. Therefore, in the limit $\eta \to 0^+$ and by means of the Dirac identity for $\delta$-distributions $\text{Im}(\lim_{\eta \to 0^+} (x+i\eta)^{-1})= -\pi \delta(x)$, the non-interacting SPSF in real-space reduces to the LDOS\,\cite{Odashima2017}, i.e.,  
\begin{align}
    A_i(\omega) = \sum_\alpha |\psi_{\alpha,i}|^2 \delta(\omega-E_\alpha) \label{ldos},
\end{align}
where $E_\alpha$ corresponds to the energy of the single-particle orbital $\psi_{\alpha,i}$ located at site $i$.

The onset of interactions will, in general, alter the SPSF in two ways. Firstly, a change in charge density through insertion or removal of an electron causes excitations in the electronic environment which change the energy of the electron and lead to a shift of the peak positions in the SPSF. 
The second effect of interactions is the emergence of satellite peaks and side-bands. In a correlated system, the spectral weight of a single-particle excitation is no longer concentrated on a single eigenstate but is spread over a range of eigenstates around the shifted energy\,\cite{Altland}. In the thermodynamic limit (and in practice for larger system or gap sizes) this distribution can become a dense continuum, effectively broadening the initial single-particle peak. The latter is associated with a decrease in the excitation's lifetime. In other words the SPSF carries information about the degree to which quasiparticle physics are present in the excitation spectrum of a correlated system, thereby indicating its amenability to single-electron measurement techniques like STS.

\bibliographystyle{prsty.bst} %abbrvnat.bst} %plainnat.bst} unsrt%}
\bibliography{Paper-References}  % needs bibtex file empty-paper.bib

\end{document}